\documentclass[a4paper,12pt]{article}

\usepackage[utf8]{inputenc}
\usepackage{fontawesome5}

\usepackage{amsmath, amssymb, amsthm}
\usepackage{bm, bbm}
\usepackage{mathtools}
\usepackage{stmaryrd}
\usepackage{siunitx}

\usepackage{graphicx}
\usepackage{xcolor}
\usepackage{tikz}
\usetikzlibrary{positioning}

\usepackage{array}
\usepackage{booktabs}
\usepackage{multirow}

\usepackage{appendix}

\usepackage{natbib}

\usepackage{algorithm, algorithmic}
\usepackage{enumerate}
\usepackage{float}
\usepackage{textcomp}
\usepackage{verbatim}
\usepackage{soul} 
\usepackage{authblk}

\usepackage{hyperref}

\usepackage[lmargin=2.4cm, rmargin=2.25cm, bmargin=3cm, tmargin=2.5cm]{geometry}
\newcommand{\mycomment}[1]{}
\newcommand\given[1][]{\:#1\vert\:}

\newcommand{\appropto}{\mathrel{\vcenter{
  \offinterlineskip\halign{\hfil$##$\cr
    \propto\cr\noalign{\kern2pt}\sim\cr\noalign{\kern-2pt}}}}}
\def\bs#1{{#1}}

\usepackage[textwidth=1.8cm, textsize=scriptsize]{todonotes}
\newcommand{\rc}{\textcolor{red}}

\newcommand{\br}{\textcolor{brown}}

\makeatletter
\newcommand{\github}[1]{%
   \href{#1}{\faGithubSquare}%
}
\makeatother

\usepackage{printlen}
\uselengthunit{mm}

\usepackage[utf8]{inputenc}
\usepackage{amsmath}
\usepackage{graphicx}
\usepackage{natbib}

\usepackage{amssymb}
\usepackage{amsthm}
\usepackage{textcomp}
\usepackage{enumerate}
\usepackage{array}
\usepackage{verbatim}
\usepackage{hyperref}
\usepackage{booktabs}
\usepackage{breqn}
\usepackage{stmaryrd} 
\usepackage{xcolor}
\hypersetup{
  colorlinks,
  linkcolor={red!50!black},
  citecolor={blue!50!black},
  urlcolor={blue!80!black}
}
\usepackage[lmargin=2.4cm,rmargin=2.25cm,bmargin=3cm,tmargin=2.5cm]{geometry}
\usepackage{ulem} 
\usepackage{xparse}
\def\bs#1{{#1}}
\newcommand{\parm}{{\theta}}

\newcommand{\rd}{{\rm d}}
\newcommand{\Obs}{{Y}^{\rm o}}

\newcommand{\obs}{{y}^{\rm o}}
\newcommand{\X}{{X}}

\newcommand{\x}{{x}}

\newcommand{\ptran}{p_{\parm}}
\newcommand{\Plik}{P_{\parm}}

\usepackage{soul} 

\usepackage{algorithmic} 
\usepackage{algorithm}
\usepackage{bbm}
\usepackage{float}
\usepackage{mathtools}
\usepackage{tikz}
\usetikzlibrary{shapes.geometric, calc} 

\usepackage{bm}
\usepackage{ulem} 

\def\bs#1{{#1}}
\begin{document}

\title{Simulation-based inference using splitting schemes 
for partially observed diffusions in chemical reaction networks}
\author[1]{Petar Jovanovski}
\author[2]{Andrew Golightly}
\author[1]{Umberto Picchini}
\author[3]{Massimiliano Tamborrino}
\affil[1]{\small Department of Mathematical Sciences, Chalmers University of Technology and the University of Gothenburg, Sweden.}
\affil[2]{\small Department of Mathematical Sciences, Durham University, UK.}
\affil[3]{\small Department of Statistics, University of Warwick, UK.}

\date{}

\maketitle

\begin{abstract}
We address the problem of simulation and parameter inference for chemical reaction networks described by the chemical Langevin equation, a stochastic differential equation (SDE) representation of the dynamics of the chemical species. This is particularly challenging for two main reasons. First, the (multi-dimensional) SDEs cannot be explicitly solved and are driven by multiplicative and non-commutative noise, requiring the development of advanced numerical schemes for their approximation and simulation. Second, not all components of the SDEs are directly observed, as the available discrete-time data are typically incomplete and/or perturbed with measurement error. We tackle these issues via \textcolor{black}{three} key contributions. First, we show that these models can be rewritten as perturbed conditionally Cox-Ingersoll-Ross-type SDEs, i.e., each coordinate, conditioned on all other coordinates being fixed, follows an SDE with linear drift and square root diffusion coefficient perturbed by additional Brownian motions. Second, for this class of SDEs, 
we develop a numerical splitting scheme  
that preserves structural properties \textcolor{black}{of the model,} such as oscillations, state space and invariant distributions, unlike the \textcolor{black}{commonly used} Euler-Maruyama scheme.  In particular, our numerical method is robust for large integration time steps. \textcolor{black}{Third,}  
we propose a sequential Monte Carlo approximate Bayesian computation 
algorithm incorporating ``data-conditional'' simulation and sequential learning of summary statistics, allowing inference for multidimensional partially observed systems, further developing previous results on fully observed systems based on the Euler-Maruyama scheme.
We validate our approach  on several models of interest in chemical reaction networks, such as the stochastic Repressilator, Lotka-Volterra, and two-pool systems, demonstrating its effectiveness, in terms of both numerical and inferential accuracy, and reduced computational cost.
\end{abstract}

\section{Introduction}

Chemical reaction networks (CRNs) play a fundamental role in the study of cellular processes, providing a framework to model and understand the stochastic behavior inherent in gene regulation, protein synthesis, and chemical signaling \citep{wilkinson2018stochastic, schnoerr2017approximation}. Proteins, which are essential biomolecules synthesized within cells, are often regulated through complex pathways that involve stochastic interactions among small populations of molecules. These interactions are influenced by both intrinsic noise, arising from the probabilistic nature of molecular reactions, and extrinsic noise, stemming from environmental fluctuations \citep{arkin1998stochastic, mcadams1997stochastic, feinberg2014epigenetic, gupta2011stochastic, paulsson2000stochastic, tian2006stochastic}. Time-course data, such as measurements of protein concentrations or gene expression levels, offer a window into these processes. Such data are typically obtained through fluorescence microscopy or other high-resolution experimental techniques, often at discrete time points and subject to measurement error \citep{young2012measuring, bar2012studying, locke2009using}. Additionally, not all components of a system are directly observed, and this partial observability, combined with measurement noise, poses significant challenges for inferring relevant biological information \citep{golightly2008bayesian, komorowski2009bayesian, bronstein2015bayesian, stathopoulos2013markov}.

CRNs are commonly modeled via Markov jump processes (MJPs), whose probability distribution over discrete molecular states satisfies the chemical master equation (CME) \citep{van1992stochastic}. While the CME provides a detailed and rigorous framework, its high-dimensional state space often renders it computationally intractable. Moreover, while it is possible to simulate exact sample paths using the stochastic simulation algorithm \citep{gillespie1977exact}, this approach becomes prohibitively expensive for larger systems due to the need to simulate every individual reaction event \citep{gillespie2013perspective}. The computational demands of inference methods that rely on this description pose a significant challenge for large-scale reaction networks. \textcolor{black}{To address this, a diffusion approximation, aiming in replacing the discrete dynamics with continuous-state approximation for systems with large molecular populations and frequent reaction events, has been proposed \citep{wilkinson2018stochastic,golightly2011bayesian}.} 
The resulting stochastic differential equation (SDE), known as the chemical Langevin equation (CLE), captures the intrinsic stochasticity of the \textcolor{black}{biochemical} system while significantly reducing its computational complexity. 
 However, the absence of closed-form solutions for 
 the SDE
 transition densities presents challenges for parameter inference, necessitating simulation-based methods for statistical estimation \citep{wilkinson2018stochastic, fuchs2013inference}, or further approximations to give a tractable process \citep{fearnhead2014inference, stathopoulos2013markov, finkenstadt2013quantifying, komorowski2009bayesian}. \textcolor{black}{Here, we focus on the former.}

\textcolor{black}{Among simulation-based inference methods \textcolor{black}{(reviewed e.g. in \citealp{cranmer2020frontier,pesonen2023abc})}, we focus here on Approximate Bayesian Computation (ABC, \citealp{sisson2018handbook}), which is suitable for parameter inference in models where forward simulation is feasible, but the likelihood function is computationally intractable, which is the case for the CLE.}
By generating \textcolor{black}{simulated data from the CLE}
and comparing it to observed data, ABC circumvents the need for closed-form transition densities, and thus likelihoods. 
However, 
its efficiency critically depends on the ability to generate accurate and computationally efficient sample paths from the CLE, placing significant emphasis on the choice of numerical solvers used for simulation.
The Euler-Maruyama (EuM) method is arguably 
the simplest \textcolor{black}{and most commonly used} scheme for the numerical simulation of CLEs (and SDEs in general), and does not rely on any specific assumptions about the structure of the CRN. 
However, it has serious shortcomings, \textcolor{black}{as it may not converge for non-globally Lipscthiz coefficients (as it often the case for CLEs), and may not preserve structural properties of the system, such as oscillations, boundary domain, hypoellipticity, even when using a very small time step, see e.g. \cite{buckwar2022splitting, kelly2023adaptive,Tubikanecetal2022}.} 

In this paper, we present \textcolor{black}{three} main contributions. The first introduces the so-called \textit{perturbed conditionally-Cox--Ingersoll--Ross (CIR)-type} SDEs, a new class of SDEs where each component, conditioned on all other coordinates being fixed, follows a CIR process (i.e. an SDE with linear drift and square-root diffusion coefficient) perturbed by additional Brownian motions. This class of models includes several important models as special cases,
such as the Lotka-Volterra model \citep{golightly2011bayesian}, the Repressilator model for oscillatory dynamics in gene expression \citep{elowitz2000synthetic}, and the two-pool model for the description of the decay and transfer of a substance between two pools \citep{nestel1969distribution}, to mention a few. The second contribution is the
the development of a novel splitting-scheme for the numerical solution of such SDEs, 
by further developing the ideas for conditionally linear systems of ODEs introduced in \cite{chen2020structure}. 
We run extensive simulation studies, showing how the proposed numerical method offers greater stability than EuM \textcolor{black}{across all parameter values} while also preserving key structural properties, such as the invariant distribution of the process and its oscillatory behavior, even at larger integration time steps. 
\textcolor{black}{This is particularly crucial when embedding this scheme into ABC algorithms: on the one hand, it allows us to choose larger simulation time steps (instead of simulating with very small ones and then subsample according to the observation time step) while still preserving the model properties
, notably reducing the computational cost. On the other hand, its numerical stability across the entire parameter space (where the model is defined) guarantees that synthetic data within ABC can always be generated, even conditionally on parameters selected from regions outside the high posterior density area.} 
Our third \textcolor{black}{contribution is on the inferential side, where we propose} an ABC sequential Monte Carlo (ABC-SMC) 
algorithm \textcolor{black}{for estimating the model parameters from partially and noisily observed  CLEs, an important problem which has been previously studied for other SDEs, see e.g. \cite{buckwar2020spectral,DitlevsenJRNMM2025,picchini2014inference,Samsonetal2025,jovanovski2023towards}. In
}  
\cite{jovanovski2023towards}, the authors constructed an ABC-SMC algorithm where the simulation of SDE solutions used a backward-smoothing technique called ``data-conditional'', taking advantage of observed data rather than forward-simulating the solutions in a ``myopic'' fashion. However, their method 
was specifically developed for univariate or fully observed bivariate SDEs \textit{without} measurement noise,  using EuM as numerical scheme. In this work, we
\textcolor{black}{further generalize that approach to}
partially observed multivariate SDEs \textit{with} measurement noise. 
 Our proposed ABC-SMC algorithm, which we call \textit{splitting-based data-conditional ABC-SMC scheme}, uses the proposed structure-preserving splitting scheme for numerical simulation, combined with data-conditional simulation and the sequential learning of automatically constructed summary statistics for ABC. This framework allows for accurate simulation and inference of relevant model parameters in multivariate SDEs. We illustrate the effectiveness of our framework 
 \textcolor{black}{on the biomathematical models mentioned above}, but the scope of the proposed inference scheme is much broader, as it can be applied to any partially observed perturbed conditionally-CIR-type SDE. 
In simulation studies, we find that the proposed ABC-SMC algorithm significantly accelerates inference, by rapidly identifying regions of high posterior density already in the initial ``rounds'' of ABC-SMC.

\textcolor{black}{The paper is organized as follows. In Section \ref{Section2}, we introduce the models of interest and the necessary background, showing that the CLEs can be rewritten as what we define as perturbed conditionally-CIR-type models. In Section \ref{sec:splitting}, we review different numerical schemes for the simulation of the one-dimensional CIR model, and then propose a novel numerical splitting scheme for perturbed conditionally-CIR-type SDEs. In Section \ref{sec:inf-problem}, we review several existing ABC algorithms before presenting our proposed splitting-based data-conditional ABC-SMC scheme in Section
\ref{sec:dc-sim}. The algorithm is tested on several biomathematical models in Section \ref{sec:simulation-studies}. A discussion follows in Section \ref{Section7}.}
\mycomment{

One of the earliest attempts [233] relied on the chemical Langevin equation approximation to the CME (section 4.1); this provides a more efficient inference scheme compared
to the auxiliary variable approach of [231], however the computational costs remain high due
to the need to compute transition probabilities for non-linear diffusion processes. 

Typically, time course data are derived from time-lapse microscopy images and fluorescent
reporters [44, 69, 148]. Advances in microscopy and fluorescent technologies are enabling
intracellular processes to be inspected at unprecedented resolutions [8,23,83,115]. Despite
these advances, the resulting data never provide complete observations since: (i) the
number of chemical species that may be observed concurrently is relatively low [148];
(ii) two chemical species might be indistinguishable from each other [59]; and (iii) the
relationships between fluorescence levels and actual chemical species copy numbers may
not be direct, in particular, the degradation of a protein may be more rapid than that of
the fluorescent reporter [69, 139]. That is, inferential methods must be able to deal with
uncertainty in observations

In order for the modelling framework to be of practical use, plausible parameter values must be obtained given data at
discrete times, that may be incomplete (in the sense of information on a subset of species in the reaction network) and
subject to error. This setting, when combined with either the MJP or CLE modelling framework precludes straightforward
likelihood-based inference owing to the intractability of the observed data likelihood.

In particular, stochastic models are often more realistic descriptions, compared with their deterministic counterparts, of many
biochemical processes that are naturally affected by extrinsic and intrinsic noise (Kærn
et al., 2005; Raj and van Oudenaarden, 2008), such as the biochemical reaction pathways that regulate gene expression (Paulsson et al., 2000; Tian and Burrage, 2006)

Given
the ubiquity of stochastic processes for real applications it is essential that eﬃcient methods are developed to enable the
analysis of modern high resolution data sets without sacriﬁcing accuracy.

An important application of stochastic processes occurs in the study of cellular processes [4,5]. Here, stochastic models of
biochemical reaction networks often provide a more accurate description of system dynamics than deterministic models [6].
This is largely due to intrinsic noise in the system dynamics

Thus, the computational
cost of evaluating Equation (1) depends on the eﬃciency of the stochastic simulation algorithm and the posterior sampler
used as a basis for likelihood-free inference. For example, ABC can be implemented using either rejection sampling

For partially observed Markov processes, pointwise evaluation of the likelihood requires the solution to the forward
Kolmogorov equation, which must be computed approximately. Therefore, standard Bayesian tools cannot be applied and
likelihood-free methods are needed, such as approximate Bayesian computation (ABC) [20–22], pseudo-marginal meth-
ods [23–25], and Bayesian synthetic likelihood (BSL) [26–28].
}

\section{Stochastic chemical kinetics}\label{Section2}
Biological processes involve complex molecular interactions, but mathematical models often simplify these by representing key stages, such as transcription, as a single chemical reaction \citep{fuchs2013inference, wilkinson2018stochastic}. A set of chemical species that interact via such chemical reactions is called a CRN. More formally, a CRN consists of a set of $d$ chemical species, $X_1, \ldots, X_d$, $X_i\in\mathbb{N}$, that interact via a network of $r$ reactions
\begin{eqnarray*}
    \nu_{1,1}^- X_1 + \nu_{2,1}^- X_2 + \ldots + \nu_{d,1}^- X_d & \xrightarrow{k_1} &\nu_{1,1}^+ X_1 + \nu_{2,1}^+ X_2 + \ldots + \nu_{d,1}^+ X_d \nonumber \\
    \nu_{1,2}^- X_1 + \nu_{2,2}^- X_2 + \ldots + \nu_{d,2}^- X_d & \xrightarrow{k_2} &\nu_{1,2}^+ X_1 + \nu_{2,2}^+ X_2 + \ldots + \nu_{d,2}^+ X_d \nonumber \\
    & \vdots &\nonumber \\
    \nu_{1,r}^- X_1 + \nu_{2,r}^- X_2 + \ldots + \nu_{d,r}^- X_d & \xrightarrow{k_r} &\nu_{1,r}^+ X_1 + \nu_{2,r}^+ X_2 + \ldots + \nu_{d,r}^+ X_d \nonumber
\end{eqnarray*}
where the stoichiometric coefficients $\nu_{i,j}^-$ and $\nu_{i,j}^+, i=1,\ldots, d, j=1,\ldots, r$ are the non-negative integer numbers of reactant and product molecules, respectively, and $k_i\textcolor{black}{> 0}$ are the kinetic rate parameters. Commonly modeled by deterministic rate equations under the law of mass action, which provide a good approximation when molecule numbers are high, CRNs may instead require stochastic models when low molecule counts make fluctuations significant \citep{elowitz2002stochastic}. 

Under well-mixed conditions and assuming thermal equilibrium, the dynamics of a chemical reaction system in a closed compartment of volume $\Omega$ depend only on molecule counts \citep{schnoerr2017approximation, fuchs2013inference}. With these assumptions, \cite{gillespie1992rigorous} derived the CME, whose solution gives the transition probability governing the continuous-time, discrete-valued Markov process, known as a MJP. For simplicity, we will use $X_j$ to denote the count of species  \textcolor{black}{$j$}, avoiding the need for a separate variable. It can be shown that the probability of the $j$-th reaction occurring in an infinitesimal time step $\rd t$ is $a_j(X) \, \rd t$, where $a_j(X)$ is the reaction's propensity function, proportional to the combinations of reactant molecules in $X = (X_1, \ldots, X_d)$. Such propensity functions take the form
\begin{equation*}
    a_j(X) = \theta_j \prod_{i=1}^d \binom{X_i}{\nu_{ij}^{-}}.
\end{equation*}
Propensity functions of this form are called mass-action kinetics type \citep{van1992stochastic}. Other types of propensity functions are also useful for various processes, such as Michaelis-Menten or Hill functions. These non-mass-action functions typically act as effective reactions, replacing several underlying microscopic reactions. These different types will be considered in the experimental section.

Denote by $X(t) = (X_1(t), ..., X_d(t))$ the state of the system (i.e., the count of species) at time $t$. 
The dynamics of this stochastic  system are governed by the CME. Let $\ptran(X(t) \given X(s))$ denote the transition probability mass function representing the probability of the system being in a particular state at time $t$, given its state at an earlier time $s<t$. The transition probability is then given by the solution to the CME
\begin{equation}
    \frac{\rd \ptran(X(t) \given X(s))}{\rd t} = \sum_{j=1}^{r} a_j(X(t) - {\nu}_j) \ptran(X(t) - {\nu}_j \given X(s)) - a_j(X(t)) \ptran(X(t) \given X(s)), \label{eq:cme}
\end{equation}
where $\nu_j = (\nu_{1,j},\ldots, \nu_{r,j})^T$ is the $j$th column of the stoichiometry matrix ${\nu}$, with elements $\nu_{i,j} = \nu_{i,j}^+ - \nu_{i,j}^-$ \citep{wilkinson2018stochastic}. Analytical solutions to the CME are only known for a limited class of systems and a few special cases. Consequently, extensive research has focused on  approximation methods. In the following, we will examine one of them, the CLE.
\subsection{Diffusion approximation: chemical Langevin equation}
The CLE and the associated Fokker–Planck equation provide a diffusion approximation to the CME, in the sense that the CLE approximates the underlying MJP \textcolor{black}{with a continuous process which solves an SDE.}
The CLE can be derived in a number of more or less formal ways. For instance, truncating terms in a Taylor expansion of the CME yields the Fokker–Planck equation, a partial differential equation describing the evolution of the probability density of a diffusion process 
(and thus a continuous-time, continuous-valued Markov process rather than a discrete-valued one; see \citealp{kramers1940brownian, moyal1949stochastic}), governed by an underlying SDE. 
For an intuitive derivation of this SDE, we refer the reader to \cite{golightly2011bayesian}. The stochastic dynamics of the $i$th chemical species $X_i$ under the CLE are given by the scalar SDE
\begin{equation}
    \rd X_i(t) = \sum_{j=1}^r \nu_{i, j} a_j(X(t)) \, \rd t + \sum_{j=1}^r \nu_{i, j} \sqrt{a_j(X(t))} \, \rd W_j(t), \quad X_i(0)=x_{i}(0), \quad i=1,\ldots, d, \label{eq:cle}
\end{equation}
where $a_j(\cdot)$ is the propensity function for reaction $j$, typically computed via a mass-action rate law, $W_j(t)$, $j=1,\ldots,r$, are uncorrelated univariate Brownian motions, and $\x_i(0)\in\mathbb{R}_+$ \textcolor{black}{is the initial count of species $i$}. For a bounded diffusion coefficient and an initial state 
\textcolor{black}{$x_0=\left(x_1(0),\ldots, x_d(0)\right)$}, there is a unique solution to \eqref{eq:cle}, 
remaining in $\mathbb{R}_+^d$ with probability one \citep{mao2006stochastic}. Like the CME, the CLE generally lacks explicit analytical solutions for most systems. However, CLE simulations are computationally more efficient 
than CME simulations, as their cost scales with the number of species $d$ rather than the frequency of reaction events, as it happens for the MJPs \citep{gillespie1977exact}. 

All species entering into the CME and the CLE \eqref{eq:cle} have a non-negative count, i.e., $X_i{(t)}\geq 0, t\geq 0, i=1,\ldots, d$, so zero is not an exit boundary, as it may be attained but not crossed. 
This may not be the case when solving the CLE numerically though, as the 
time discretization of the CLE may yield square roots of negative values, resulting in an ill-defined process \citep{anderson2019constrained, wilkie2008positivity, szpruch2010comparing, dana2011physically, schnoerr2014complex}. This is the case, for example, when considering the commonly used EuM method, which motivates us to derive an alternative boundary-preserving numerical scheme; see Section \ref{sec:splitting}.
Negative values in the EuM simulations are handled by either truncating at zero or by taking their absolute values. However, this introduces a bias in the model dynamics, whose quantification and impact on the inference is not easily quantifiable, as discussed in Section \ref{sec:simulation-studies}.

\subsection{CLEs as \textcolor{black}{perturbed} conditionally-CIR-type models}\label{condCIR}
Many deterministic CRNs are conditionally linear, i.e., the resulting ordinary differential equation (ODE) associated with the $i$th component is linear in $X_i$ when all other $X_j, j\neq i$ are fixed (see Appendix \ref{CLODE} for a description of a numerical splitting scheme for conditionally linear ODEs, and an illustration for the deterministic Repressilator). However, this is not the case for the stochastics CRNs given by the CLE \eqref{eq:cle}, as the the driving noise is of square-root type. Instead, the CLE \eqref{eq:cle} has what we call a \textit{\textcolor{black}{perturbed} conditionally-CIR-type} structure. To define what we mean by this, we start by recalling the one-dimensional CIR process, which is given by the It{\^o} SDE
\begin{equation}
    \rd X(t) = \beta (\alpha - X(t)) \, \rd t + \sigma \sqrt{X(t)} \, \rd W(t), \label{eq:cir}
\end{equation}
with initial state $X(0) = x_0  > 0$, where $W(t)$ is a one-dimensional Brownian motion, and $\alpha, \beta, \sigma$ are positive parameters. The solution to the CIR process \eqref{eq:cir}, which is not known in closed form, is almost surely non-negative, and if the Feller condition $2 \alpha \beta > \sigma^2$ is satisfied, it is almost surely positive \citep{feller1951two, feller1951diffusion, cox2005theory}. 

For many biochemical reaction networks, we notice that the $i$th chemical species $X_i$ under the CLE \eqref{eq:cle} can be rewritten as
\begin{equation}
\begin{aligned}
    \rd X_i(t) & = \sum_{j \in R_{-i}} \nu_{i,j} a_j(X_{-i}(t)) \, \rd t + \sum_{j \in R_i} \nu_{i,j} a_j(X_{-i}(t))X_i(t) \, \rd t \\ 
    & + \sum_{j \in R_{-i}} \nu_{i,j} \sqrt{a_j(X_{-i}(t))} \, \rd W_{j}(t) + \sum_{j \in R_i} \nu_{i,j} \sqrt{a_j(X_{-i}(t))X_i(t)} \, \rd W_{j}(t),\quad i=1,\ldots, d
\end{aligned} \label{eq:cle-componentsde}
\end{equation}
by explicitly separating the terms depending on $X_i$ from those that do not depend on it. In particular, the index set $R_i \subseteq \{1, 2, ..., r\}$ denotes the set of reactions involving the species $X_i$, whereas $R_{-i}$ includes those reactions that do not involve it. Similarly, the vector $X_{-i}(t)$ represents the concentrations of all species except $X_i$, that is $X_{-i}(t)=(X_1(t), \ldots, X_{i-1}(t), X_{i+1}(t), \ldots, X_d(t))$. The propensity functions are adjusted based on the involvement of $X_i$: if $j \in R_i$, then $a_j(X_{-i}(t))$ is computed for all species except $X_i$, and is multiplied by $X_i(t)$ to account for mass-action kinetics. Conversely, for reactions in $R_{-i}$,  $a_j(X_{-i}(t))$ does not include $X_i(t)$ by definition.

Hence, conditioned on all components $X_{-i}$ being fixed, the SDE \eqref{eq:cle-componentsde} for species $i$ can be rewritten as 
\begin{equation}
    \rd X_i(t) = \left(\widetilde{a}_{i} - \widetilde{b}_{i} X_i(t) \right) \, \rd t + \left(\sum_{j \in R_{i}} \widetilde{c}_{i, j} \, \rd W_{j}(t)\right) \sqrt{X_i(t)}+ \sum_{j \in R_{-i}} \widetilde{c}_{i, j} \, \rd W_{j}(t),\quad i=1,\ldots, d\label{eq:cle-cond-gen}
\end{equation}
where $\widetilde{a}_{i}, \widetilde{b}_{i}, \widetilde{c}_{i, j}$ are constants given by
\begin{equation}
    \widetilde{a}_{i} = \sum_{j \in R_{-i}} \nu_{i,j} a_j(X_{-i}(t)),\qquad \, \widetilde{b}_{i} = -\sum_{j \in R_i} \nu_{i,j} a_j(X_{-i}(t)), \qquad\, \widetilde{c}_{i, j} = \nu_{i,j} \sqrt{a_j(X_{-i}(t))}. \label{constant} 
\end{equation}
From \eqref{eq:cle-cond-gen}, we see that, conditionally on all components $X_{-i}$ being fixed, the SDE of the $i$th species looks like that of the CIR process \eqref{eq:cir} with $\beta=\widetilde{b}_i$, $\alpha=\widetilde a_i/\widetilde{b}_i$ and $\sigma=\sqrt{\sum_{j\in R_i}\widetilde c_{i,j}^2}$, perturbed by an extra Brownian component given by $\sum_{j\in R_{-i}}\widetilde c_{i,j}dW_j(t)$. This is why we call \eqref{eq:cle-componentsde} a \textit{\textcolor{black}{perturbed}} conditionally-CIR-type process, as conditionally on the other components $X_{-i}$ being fixed, the SDE for species $i$ {resembles} that of the CIR process \textcolor{black}{perturbed by some Brownian noise}.

\subsubsection{Examples \textcolor{black}{and features} of perturbed conditionally-CIR models}
Well known models that belong to our perturbed conditionally-CIR-type framework include the stochastic versions of the Lotka-Volterra,  Michaelis-Menten enzyme kinetics, the Repressilator model, which describes oscillatory dynamics in gene expression, and the two-pool model, which explains the decay and transfer of a substance between two pools. This fit is exemplified in Section \ref{sec:simulation-studies}. 
However, if the SDE for species \textcolor{black}{$i$} incorporates a propensity function that is nonlinear in $X_i$, such as in the Schl{\"o}gl model \citep{schlogl1972chemical}, then it cannot be represented in the form specified in \eqref{eq:cle-componentsde}. 

While in \textcolor{black}{perturbed} conditionally-CIR-type form, the SDE for species $i$ resembles that of the CIR process, there are some notable differences \textcolor{black}{between the two. For example,} sometimes the sign before $X(t)$ is positive \textcolor{black}{ in \eqref{eq:cle-cond-gen}}, unlike the negative sign in \eqref{eq:cir}. Moreover, in certain models, such as the stochastic Lotka--Volterra system, the condition ($2 \alpha \beta > \sigma^2$) ensuring almost sure positivity \textcolor{black}{of the solution in the \eqref{eq:cle-split1} formulation}  
is not satisfied for any parameter configuration. 
In other models, positivity may hold or not across the simulation timespan due to dependencies on other states of the system.
In the next section, we introduce a novel splitting scheme for solving SDEs belonging to the \textcolor{black}{perturbed} conditionally-CIR class.

\section{Splitting schemes for \textcolor{black}{perturbed} conditionally-CIR-type SDEs}
\label{sec:splitting}

Consider a discretized time interval $[0,T]$ with equidistant observation time steps $\Delta=t_l-t_{l-1}$, $l=1,\ldots, n$, $t_0=0$ and $t_n=T$. In experimental scenarios, data are often collected at low frequencies, leading to relatively large $\Delta$. Simulating the SDE solution via a (not-exact) numerical scheme with such time steps may then compromise the accuracy of the approximated solutions, and thus of the posterior inference. A common approach to mitigate this consists of simulating the SDE with a smaller time step $h=\Delta/A$, where $A\geq 2$ denotes the number of subintervals in each observational interval $[t_{{l}-1}, t_{{l}}], {l}=1,\ldots, n.$ That is, the SDE is simulated at a refined grid $\tau_0=t_0=0, \ldots, \tau_{nA}{=t_n}=T$, with $\tau_j = j h$ and $\tau_{{l}A}=t_{{l}}$,
but only the states at the observation times $t_i$ are retained for inference. 
Here, we denote by $\widetilde X_{i,k}$ the numerical solution of \eqref{eq:cle} for the $i$th species at time $\tau_k$, i.e., the numerical approximation of $X_i(\tau_k)$, by $\widetilde X_{1:d, k}$ the numerical approximation of the entire species $X(\tau_k)$, and by $\widetilde X_{1:d, 0:nA}$ the numerical approximation of $X(\tau_k)$ for $k=0,\ldots, nA$, i.e., the numerical approximation of $X(t)$ over the discretized time interval $[0,T]$, assuming $\widetilde X_{1:d, 0}=X(0)$. 
The most commonly used method for simulating the CLE \eqref{eq:cle} is the EuM scheme. The EuM approximation of the CLE of the $i$th chemical species at time $\tau_{k+1}$ starting from $ \widetilde X^{\textrm{EuM}}_{i,k}$ at time $\tau_k$, denoted by $ \widetilde X^{\textrm{EuM}}_{i,k+1}$,  
is given by 
\begin{equation}
     \widetilde X^{\textrm{EuM}}_{i, k + 1} =  \widetilde X^{\textrm{EuM}}_{i, k} + \sum_{j=1}^r \nu_{i, j} a_j( \widetilde X^{\textrm{EuM}}_{i, k}) h + \sum_{j=1}^r \nu_{i, j} \sqrt{a_j( \widetilde X^{\textrm{EuM}}_{i, k})} \Delta W_{j, k + 1}, \quad \Delta W_{j, k + 1} \sim \mathcal{N}(0, h).  \label{eq:EuM}  
\end{equation}
However, it is well known that the EuM fails to preserve some important structural properties, e.g. the non-negativity of the trajectories, as discussed in the next section for the CIR process. 
The idea behind the proposed splitting scheme is to derive a property-preserving numerical method taking advantage of the perturbed conditionally-CIR-type property of the CLE \eqref{eq:cle}, obtained, under some conditions on $a_j(X(t))$, by solving the $d$-dimensional system of CLEs componentwise in $X_i$, while keeping all other components $X_{-i}$ fixed, as discussed in Section \ref{condCIR}. Before describing our proposed splitting scheme in Section \ref{sec:cond-cir}, we briefly revise some existing numerical schemes for the one-dimensional CIR process.

\subsection{Simulation of the one-dimensional CIR process}\label{1dCIR}
Even if an explicit solution to the CIR  \eqref{eq:cir} is not available in closed form, we can simulate from it exactly, as its conditional transition density is known to follow a non-central chi-square distribution. Nevertheless, several numerical schemes have been proposed in the literature to approximate its solution, \textcolor{black}{as it is often considered as an important model to \lq\lq test\rq\rq new numerical methods} due to the numerical challenges it poses. Indeed, canonical schemes (such as EuM) fail to preserve the positivity of its trajectories, and the common assumptions required for weak and strong convergence results \citep{kloeden1992stochastic} do not apply due to the diffusion coefficient not being Lipschitz. This is why EuM does not converge, despite being still extensively used, typically with an added truncation at zero, or by taking absolute values to guarantee the non-negativity of the approximations.  
Modified versions of EuM or of the Milstein scheme have been proposed to tackle this, e.g. tamed EuM, truncated EuM or truncated Milstein 
\citep{higham2005convergence,lord2010comparison, cozma2020strong,hefter2018strong}. 

Here, we focus on a different approach consisting of: i) transforming the SDE with multiplicative noise \eqref{eq:cir}  into one with additive noise by applying the Lamperti transformation $Z(t) = \sqrt{X(t)}$ to \eqref{eq:cir}; ii) discretizing the resulting SDE, e.g. with an implicit or splitting scheme
\citep{alfonsi2005discretization, alfonsi2013strong, dereich2012euler, chassagneux2016explicit, kelly2023adaptive, kelly2022role}; iii) invert the Lamperti transform, mapping the numerical solution of the Lamperti-transformed SDE to the original SDE, i.e. $X(t)=Z(t)^2$. 
More precisely, the Lamperti-transformed SDE takes the form 
\begin{equation*}
    \rd Z(t) = \left(-\frac{b Z(t)}{2} + \frac{4a - \sigma^2}{8 Z(t)}\right)\, \rd t + \frac{\sigma}{2} \, \rd W(t).
\end{equation*}
The square root in the diffusion coefficient is now removed,  mitigating the issue of negative values for Taylor-based schemes (such as EuM). However, care must still be taken to prevent solutions from reaching zero due to the nonlinear term $1 / Z(t)$ appearing in the drift of the Lamperti-transformed SDE. We refer to \cite{kelly2023adaptive} and to Section 8.5.2 of \cite{kellycomputation} for a comprehensive review on these simulation techniques. In the next section, we use these methods within the proposed numerical scheme for the \textcolor{black}{perturbed} conditionally-CIR-type process.
 
\subsection{Splitting schemes for \textcolor{black}{perturbed} conditionally-CIR-type SDEs}
\label{sec:cond-cir}
Here, we derive a numerical splitting scheme for CLEs \eqref{eq:cle} which can be rewritten as \eqref{eq:cle-componentsde}. This requires three steps: 
(a) split \eqref{eq:cle-componentsde} into explicitly solvable subequations; (b) solve the subequations explicitly; (c) compose the solutions of the subequations in a suitable manner. 
In the following, we detail these three steps for the proposed numerical splitting scheme.

\subsubsection{Step (a): Choice of the subequations}\label{sec:subeq}
Conditionally on all components $X_{-i}$ being fixed, the dynamics of the $i$th species in \eqref{eq:cle-componentsde} are given by \eqref{eq:cle-cond-gen}. To solve this perturbed conditionally-CIR-type SDE componentwise, we consider \textcolor{black}{whether any of the Brownian motions in the perturbation $\sum_{j\in R_{-i}}\widetilde c_{i,j}dW_j(t)$ in equation~\eqref{eq:cle-cond-gen} are or not shared across species, yielding  the following two scenarios.} \\
\noindent
\textit{1) Brownian motions shared across species.} Brownian motions shared among chemical species cannot be combined together into a single
Brownian motion, as they introduce correlations between species. When this happens, we split the SDE \eqref{eq:cle-cond-gen} into the following two subequations:
\begin{eqnarray}
 \rd X^{(1)}_i(t)& =& \left(\widetilde{a}_{i} - \widetilde{b}_{i} X^{(1)}_i(t) \right) \, \rd t + \left(\sum_{j \in R_{i}} \widetilde{c}_{i, j} \, \rd W_{j}(t)\right) \sqrt{X^{(1)}_i(t)}, \label{eq:cle-split1}\\
 \rd X^{(2)}_i(t) &=& \sum_{j \in R_{-i}} \widetilde{c}_{i, j} \, \rd W_{j}(t), \label{eq:cle-split2}
 \end{eqnarray}
where the first is a \textcolor{black}{conditionally-}CIR-type SDE, and the second is an explicitly solvable SDE consisting of a sum of Brownian motions. \\
\noindent
\textit{2) Brownian motions not shared across species.} If the Brownian motions $W_{j}, j \in R_i \cup R_{-i}$ entering into \br{\eqref{eq:cle-cond-gen}} for the chemical species $X_i$ are absent from the SDEs of other species, their effects can be combined into a single Brownian motion process scaled by the square root of $\sum_{j \in R_{-i}} \widetilde{c}_{i, j}^2 + X_i \sum_{j \in R_i} \widetilde{c}_{i, j}^2$. Then, 
\eqref{eq:cle-cond-gen} can be rewritten as a \textcolor{black}{conditionally-}CIR-type SDE similar but not identical to \eqref{eq:cle-split1}, as the Brownian motions $W_{j}$ for $j\in R_{-i}$ are also considered.

Note that we only focus on the first and more general scenario, as the second can be similarly derived. Then, after having derived the subequations \eqref{eq:cle-split1}-\eqref{eq:cle-split2}, the next step consists in solving them.

\subsubsection{Step (b): Solution of the subequations \eqref{eq:cle-split1} and \eqref{eq:cle-split2}}
Here, we provide the solutions of the subequations \eqref{eq:cle-split1}-\eqref{eq:cle-split2} in $[\tau_k,\tau_{k+1}]$ starting from $x_{i,k}$ at time $\tau_k$. 
The solution of the second subequation \eqref{eq:cle-split2} at time $\tau_{k+1}$ with initial state $x_{i,k}$, denoted by $\Phi_h^{(2)}(x_{i, k}):=\textcolor{black}{X^{(2)}_{i} (\tau_{k+1})}$, $i=1,\ldots, d$, is given by 
\begin{equation}
    \Phi_h^{(2)}(x_{i, k}) = x_{i, k} + \sum_{j \in R_{-i}} \widetilde{c}_{i, j} \Delta W_{j, k + 1}, \quad \Delta W_{j, k + 1} \sim \mathcal{N}(0, h), \label{eq:cle-split2-sol}
\end{equation}
\textcolor{black}{where $\Delta W_{j,k+1},\ j\in {R}_{-i}$ are independent and identically distributed (iid) Gaussian increments.}
As for the first equation \eqref{eq:cle-split1}, we apply the Lamperti transform to obtain an SDE with constant diffusion, as discussed in Section \ref{1dCIR}. In particular, applying the transform $Z = f(X)=\sqrt{X}$ to \eqref{eq:cle-split1} yields the SDE
\begin{equation}
    \rd Z_i(t) = \left( -\frac{\widetilde{b}_i}{2} Z_i(t) + \left( \frac{\widetilde{a}_i}{2} - \frac{1}{8} \sum_{j \in R_{i}} \widetilde{c}^2_{i,j}  \right) \frac{1}{Z_i(t)} \right) \, \rd t + \frac{1}{2} \sum_{j \in R_{i}} \widetilde{c}_{i, j} \, \rd W_{j}(t). \label{eq:cle-lamperti}
\end{equation}
As this SDE cannot be solved exactly, we again rely on a splitting scheme, decomposing it into the following subequations 
\begin{eqnarray}
 \rd Z^{(1a)}_i(t) &=& \left( -\frac{\widetilde{b}_i}{2} Z^{(1)}_i(t) + \left( \frac{\widetilde{a}_i}{2} - \frac{1}{8} \sum_{j \in R_{i}} \widetilde{c}^2_{i,j}  \right) \frac{1}{Z^{(1)}_i(t)} \right) \, \rd t, \label{eq:bern-ode}\\
   \, \rd Z^{(1b)}_i(t)& =&  \frac{1}{2} \sum_{j \in R_{i}} \widetilde{c}_{i, j} \, \rd W_{j}(t),\label{eq:const-sde}
   \end{eqnarray}
   where the first is a Bernoulli ODE and the second is a sum of Brownian motions with zero drift  and constant diffusion coefficients. The solution of \eqref{eq:bern-ode} \textcolor{black}{at time $\tau_{k+1}$} with initial state $\textcolor{black}{z_{i, k}}$ is given by \textcolor{black}{$\Phi_h^{(1a)}(z^{(1)}_{i, k}):=\textcolor{black}{Z}^{(1a)}_{i}(\tau_{k+1})$}, where
\begin{equation}
    \Phi_h^{(1a)}(\textcolor{black}{z}_{i, k}) = \sqrt{4 \widetilde{a}_i - \sum_{j \in R_{i}} \widetilde{c}^2_{i,j} + e^{-\widetilde{b}_i h} \left(-4 \widetilde{a}_i + \sum_{j \in R_{i}} \widetilde{c}^2_{i,j} + 4 \widetilde{b}_i (\textcolor{black}{z}_{i, k})^2\right)} \left/ 2 \sqrt{\widetilde{b}_i} \right.. \label{eq:ode-sol}
\end{equation}
Note that, the solution of the ODE might become complex for some values of the coefficients $\widetilde a_i, \widetilde b_i, \widetilde c_{i,j}$. One could choose an adaptive 
time step $h$ to prevent this from happening. 
However, we opt to not use adaptive time stepping techniques but, instead, 
set the solution to zero if it is complex. 
On the one hand, this introduces a bias, as we know that the true solution is positive. On the other hand, this guarantees the state-space preservation without the additional \lq\lq complexity\rq\rq of the time-adaptive steps. The impact of this choice, as well as alternative splitting decompositions for \eqref{eq:cle-lamperti}, remains an area for further investigation. 

The solution to \eqref{eq:const-sde} at time $\tau_{k+1}$ with initial state \textcolor{black}{$z_{i, k}$}, denoted by \textcolor{black}{$\Phi_h^{(1b)}(z_{i, k})=Z^{(1b)}_{i}(\tau_{k+1})$} is given by 
\begin{equation}
\Phi_h^{(1b)}(\textcolor{black}{z}_{i, k}) = z_{i, k} + \frac{1}{2} \sum_{j \in R_{i}} \widetilde{c}_{i, j} \Delta W_{j, k + 1}, \quad \Delta W_{j, k + 1} \sim \mathcal{N}(0, h). \label{eq:linear-sol}
\end{equation}
The numerical solution of the Lamperti-transformed SDE \eqref{eq:cle-lamperti} at time $\tau_{k+1}$ starting from $z_{i,k}$ is then obtained with the Lie-Trotter composition as $\widetilde Z_{i,k+1}=(\Phi_h^{(1b)} \circ \Phi_h^{(1a)})(z_{i,k})$. The inverse map $f^{-1}(z)=z^2$ is then used to obtain the numerical solution of \eqref{eq:cle-split1} at time $\tau_{k+1}$ starting from $x_{i,k}$ as
\begin{equation}\label{solBernoulli}
    \widetilde X^{(1)}_{i, k + 1} = \left((\Phi_h^{(1b)} \circ \Phi_h^{(1a)})(\sqrt{x_{i, k}})\right)^2.
\end{equation}

\subsubsection{Step (c): Composition of the solutions}
\textcolor{black}{After having solved subequations \eqref{eq:cle-split1} and \eqref{eq:cle-split2} numerically and explicitly, respectively, we obtain the componentwise solution of the $i$th species at time $\tau_{k+1}$ starting from $x_{i,k}$ via the Lie-Trotter composition of the solutions \eqref{solBernoulli} and \eqref{eq:cle-split2-sol} to give
}
\begin{equation}\label{solSDE}
    \widetilde X_{i, k + 1} = \left((\Phi_h^{(1b)} \circ \Phi_h^{(1a)} \circ \Phi_h^{(2)})(x_{i, k})\right)^2.
\end{equation}
This step is positivity-preserving, with the caveat that the square root of the solution to the nonlinear ODE may become complex. 
\textcolor{black}{Hence, we have obtained the updated $i$th species when all other $j\neq i$ components are assumed fixed. We can then proceed iteratively to update them componentwise, always using the latest updated components. The order in which the components are  updated, as well as the composition method, depend on the specific process, as shown in Section \ref{sec:simulation-studies}.}

\section{Bayesian inference}
\label{sec:inf-problem}
In this section, we consider the inference problem: given noisy and/or partial observations of the system, our goal is to compute the posterior distribution of the parameter vector $\theta = (k_1, k_2, \dots, k_{r}, \lambda)$, including both  the kinetic rate constants $k_j$ for each reaction $j$, but also additional elements $\lambda\textcolor{black}{ \in\Lambda\subseteq\mathbb{R}^p,\ p\in\mathbb{N}}$, representing, for example, Hill constants and functions of the observation error, such as its variance.
In many experiments, only partial observations of the system are available, meaning that only a subset of chemical species is observed at a given time $t$. We denote by $X^{{o}}(t)$ and $X^{{u}}(t)$ the observed and unobserved states of $X(t)$ at time $t$, with dimensions $d_{{o}}$ and $d_{{u}}$, respectively, with $d=d_{{o}}+d_{{u}}$. In particular, we take $X^{{o}}(t)=L X(t)$, where, $L$ is a $d_{\rm o} \times d$ matrix that maps the state vectors to the observed components. For example, if $L = I_d$, the $d$-dimensional identity matrix, all components are observed. Thus, at any time $t$, the state space $\mathcal{X}$ of $X(t)$ is partitioned into an observed subspace $\mathcal{X}^{{o}} \subseteq \mathbb{R}^{d_o}$ and an unobserved subspace $\mathcal{X}^u \subseteq \mathbb{R}^{d_u}$,  i.e.,  $X^{{o}}(t)\in\mathcal{X}^o,  X^{{u}}(t)\in\mathcal{X}^u, X(t)\in\mathcal{X}$.

Assume that the diffusion process is observed at $n$ discrete \textcolor{black}{equidistant} time points $t_l = l \Delta$, for $l = 0, \ldots, n$.  
Denote by \textcolor{black}{$\Obs(t)$ the output process, $\obs=(\Obs(t_0),\ldots, \Obs(t_n))$ its observed sample path, and $x^\circ=(X^\circ(t_0),\ldots, X^\circ(t_n))$ and $x^u=(X^u(t_0),\ldots, X^u(t_n))$ the observed and unobserved components at the given measurement times.}
Two types of available observations are considered here: (i)
without measurement error, such that $\Obs(t_l)=X^{\textrm{o}}(t_l)=LX(t_l) \in\mathcal{X}^o$; (ii) with measurement error, such that $\Obs(t_l)=X^{\textrm{o}}(t_l)+\xi_l=LX(t_l)+\xi_l$,
with the iid terms $\xi_l$ typically modeled as 
\textcolor{black}{$d_0$-dimensional} Gaussian \textcolor{black}{distributions with mean vector $0$ and covariance matrix $\Sigma^{\rm o}$, i.e.} $\xi_l\sim\mathcal{N}_{d_{\rm o}}(0, \Sigma^{\rm o})$. Given an observed sample path  \textcolor{black}{$\obs$}
, and a prior density $\pi(\theta)$ ascribed to $\theta$, the goal is to compute the posterior distribution of $\theta$:
\begin{equation}
    \pi(\theta \given y^{\textrm{o}}) \propto \pi(\theta) p_\theta(y^{\textrm{o}}), 
    \label{eq:bayes-rule}
\end{equation}
\textcolor{black}{where $p_\theta(\obs)$ denotes the likelihood function.}

Depending on the available type of observations, the likelihood 
can take different forms. If the observations are exact (without measurement error, so $y^{\textrm{o}}=x^{\textrm{o}}$), using the Markov property of the model, the likelihood function takes the form
\begin{equation}
    p_\theta(y^{\textrm{o}}) =
    \int_{(\mathcal{X}^u)^n} \prod_{l=1}^n \ptran(x^{\textrm{o}}(t_l), x^{\textrm{u}}(t_l) \given x^{\textrm{o}}(t_{l-1}), x^{\textrm{u}}(t_{l - 1})) \, \rd x^{\textrm{u}}(t_l), \label{eq:exact-obs}
\end{equation}
where \textcolor{black}{$p_{\theta}(x(t_l) \mid x(t_{l-1}))=p_\theta(x^{{o}}(t_l),x^{{u}}(t_l) \mid x^{{o}}(t_{l-1}),x^{{u}}(t_{l-1}))$ 
} denotes the transition density of the process
$X$ 
from $X(t_{l-1})=x(t_{l-1})$ at time $t_{l-1}$ to $X(t_{l})=x(t_l)$ at time $t_{l}$,  
and the integral is 
\textcolor{black}{on $X^{{u}}(t_l)=x^{{u}}(t_l)$} over the unobserved subspace $(\mathcal{X}^u)^n$. If the observation model includes Gaussian measurement error $\xi_l\sim\mathcal{N}_{d_{\rm o}}(0, \Sigma^{\rm o})$ \textcolor{black}{ with probability density function (pdf) $\phi_{d_o}(\cdot \given \mu, \Sigma)$,} 
then the likelihood function takes the form
\begin{equation}
    p_\theta(y^{\textrm{o}}) = \int_{\mathcal{X}^n} \prod_{l=1}^n \phi_{d_o}(y^{\textrm{o}}(t_l) \given x(t_l), \Sigma^{\rm o}) \ptran\left(x(t_l) \given x(t_{l - 1})\right) \, \rd x(t_l). \label{eq:meas-err-obs}
\end{equation}
Note that the integral is taken over the complete latent space $\mathcal{X}^n$, regardless of how many components are observed, since the process $X^{\textrm{o}}$ is not directly observed either, but only indirectly  through noisy measurements. In either case \eqref{eq:exact-obs} or \eqref{eq:meas-err-obs}, the likelihood function does not have a closed-form expression.

\subsection{Simulation-based inference}
Bayesian inference is \textcolor{black}{often} complicated by the intractability of the likelihood function, necessitating simulation-based approaches such as particle MCMC (pMCMC) \citep{andrieu2009pseudo,andrieu2010particle}, whose most basic version requires only forward simulation of the SDE and pointwise evaluation of the observation density. However, this approach can be computationally prohibitive. Other simulation-based methods, such as ABC, also bypass the requirement for an explicit likelihood function by using a \textit{model simulator} to generate synthetic data from the model, and can provide a computationally efficient, albeit approximate, alternative to pMCMC (e.g. \citealp{wang2024comprehensive} for a review targeting computational biology). 
In the case of SDEs, the model simulator is the SDE solver, \textcolor{black}{which} is typically a numerical approximation, meaning the simulation is subject to a discretization error\textcolor{black}{, adding a further approximation to the derived posterior}.  
In particular, ABC methods for SDEs involve repeatedly sampling parameter-path pairs 
$(\parm, y)$, where parameters $\theta$ are drawn from a proposal density, and paths $y$ are simulated using a numerical solver conditioned on these draws. The simulated dataset $y$ is then generated at discrete observation time points, with or without additional measurement noise depending on the observation regime, as described in Section~\ref{sec:inf-problem}.
Parameters that generate paths $y$ closely matching the observation $\obs$ are considered to originate from a region of high posterior density. Typically, it is necessary to first summarize the data using low-dimensional summary statistics, denoted by $S(\cdot)$. Then, the ABC approximation to the posterior \eqref{eq:bayes-rule}, becomes:
\begin{equation}
\pi_\epsilon(\theta \given S(\obs))\propto \pi({\theta}) \int \mathbbm{1}( \Vert S(y) - S({\obs}) \Vert \leq \epsilon) {p_\theta}(S(y)) \, {\rm d}{y}, \label{eq:abc-posterior}
\end{equation}
where $\|\cdot\|$ is an appropriate distance metric, $\mathbbm{1}(A)$  the indicator function of the set $A$, 
and $\epsilon>0$ a tolerance value \textcolor{black}{on the distance between simulated and observed summary statistics,} determining the accuracy of the approximation. The likelihood of the summary statistics $p_\theta(S(y))$
is implicitly defined with samples obtained by first simulating a path $y$
and then computing $S(y)$. Using a small $\epsilon$ and  ``informative'' (ideally sufficient) $S(\cdot)$, ABC may produce a reasonable approximation to the true posterior $\pi(\theta \given \obs)$. However, choosing an appropriate $\epsilon$ involves a trade-off: reducing $\epsilon$ increases computational demands due to a higher rejection rate of proposed parameters, yet it potentially improves the accuracy of the posterior inference, provided a sufficient number of proposals is accepted. For high-dimensional data originating from SDEs, selecting summary statistics that retain as much information about $\theta$ is crucial. A seminal work for constructing summary statistics in a semi-automatic way is by \cite{fearnhead2012constructing}, who showed that the posterior mean serves as an optimal summary statistic under quadratic loss, and that such expectation (and hence the summary statistic) can be estimated by fitting a linear regression model to a set of prior-predictive samples. This methodology was successfully applied to SDEs in \cite{picchini2014inference}.
The linear regression approach has subsequently evolved to the use of deep neural networks \citep{jiang2017learning, wiqvist2019partially, alsing2018massive, brehmer2020mining, aakesson2021convolutional,jovanovski2023towards}. \textcolor{black}{In particular, \cite{jovanovski2023towards} focused on SDEs, considering}  
the ``partially exchangeable network'' (PEN) introduced by \cite{wiqvist2019partially} 
 to learn the summary statistics from discretely observed Markov processes. 
Although the posterior mean as a summary statistic is optimal under quadratic loss, it is only locally sufficient; globally sufficient summary statistics have been introduced by \cite{chen2020neural, chen2023learning}. An alternative way of proposing summary statistics for SDEs is by choosing them based on model properties, such as invariant density, invariant spectral density and correlations between coordinates, see e.g. \cite{buckwar2020spectral,DitlevsenJRNMM2025,Samsonetal2025}.

\subsection{ABC-SMC}

Currently, ABC-SMC is the state-of-art ABC algorithm. 
ABC-SMC samplers run through a series of $R$ iterations (``rounds''), which we index with $r$, \textcolor{black}{producing weighted samples also known as ``particles'',} from increasingly accurate ABC posterior approximations. ABC-SMC samples particles from the prior $\pi(\theta)$ at the first round ($r=1$), and \textcolor{black}{then from a iteratively refined proposal kernel over successive iterations,} 
while reducing the tolerance level $\epsilon$. This approach systematically improves the approximation to the posterior as the algorithm progresses. After the first ABC-SMC round, the approximate posterior $\pi_{\epsilon_1}$ is represented by a set of $M$ equally weighted samples ${\theta}_1^{1:M}$. 
After the first round, parameters are no longer drawn from the initial prior but are instead proposed through perturbations of the weighted particles from the previous round. Specifically, at a generic iteration $r>1$, a particle $\theta^*$ is first sampled from the particle population ${\theta}^{1:M}_{r - 1}$ using probabilities $w_{r - 1}^{1:M}$, and then perturbed using a transition kernel \textcolor{black}{$\mathcal{K}(\cdot \given \theta^*)$ which} is used to both move particles into regions of potentially high\textcolor{black}{er} posterior density and enhance \textcolor{black}{their diversity.} 
 A common choice for such kernel is a Gaussian distribution, \textcolor{black}{i.e. $\theta_r^i\sim \mathcal{N}(g(\theta^*),\Sigma_r)$, 
 with mean (function of $\theta^*$) and covariance specified in several possible ways \citep{beaumont2009adaptive, toni2009approximate,picchini2022guided}}. Since parameters are sampled from the particles of the previous round, \textcolor{black}{ following the principles of SMC sampling \citep{del2006sequential}, the approximate posterior $\pi_{\epsilon_r}$ is constructed as $\pi_{\epsilon_r}({\theta} \given S(\obs)) = \sum_{j = 1}^M w_r^j \mathcal{K}({\theta}_r^j \given {\theta}_{r - 1}^j)$, where} the importance weights for the new particle population $\theta_r^{1:M}$  are given by
\begin{equation}
    w^i_r \propto \frac{\pi({\theta}_r^i)}{\sum_{j=1}^M w_{r - 1}^j \phi_p({\theta}_r^i \given {\theta}_{r - 1}^j, {\Sigma}_{r - 1})},\qquad i=1,\dots,M. \label{eq:particle-weight-abcsmc}
\end{equation}
A particular version of ABC-SMC is found in  Algorithm \ref{algo:sl-abs-smc} in Appendix \ref{sec:appendix-dcabcsmc}, which is the one we used in our experiments when we consider \textit{data conditional} ABC-SMC, described in Section \ref{sec:dc-sim}. In this paper, \textcolor{black}{we have $g(\theta^*)=\theta^*$, and} ${\Sigma}_r$ is chosen to be twice the (weighted) covariance-matrix of the current particle population (as in \citealp{beaumont2009adaptive,filippi2013optimality}), but a more efficient and flexible class of Gaussian and copula samplers is presented in \cite{picchini2022guided}, which we do not employ here for simplicity. The sequence of thresholds $\epsilon_1 > \ldots > \epsilon_T$ is not predetermined, but dynamically adjusted during \textcolor{black}{the execution}, 
usually based on a percentile of the simulated \textcolor{black}{or accepted} distances. Regarding the determination of the summary statistics for ABC, as mentioned in the previous section, in \cite{jovanovski2023towards} the dynamic learning strategy from \cite{chen2020neural} was combined with a PEN \citep{wiqvist2019partially}, to iteratively learn the summaries for Markovian time series, while retraining PEN in each ABC-SMC round with increasingly more informative parameter-path pairs. Although the obtained summary statistics were only locally sufficient, the simulation study demonstrated that the approach yields accurate posterior inference. To accommodate multidimensional time series, we implement the initial layers in PEN as convolutional layers with $d_o$ output channels. A similar approach has been adopted in \citep{chen2020neural, aakesson2021convolutional}.

\section{Splitting-based data-conditional ABC-SMC}
\label{sec:dc-sim}
In the previous sections, we introduced a novel splitting scheme for \textcolor{black}{perturbed conditionally-}CIR-type SDE models and discussed simulation-based inference methods, with a focus on ABC-SMC. We now build on the data-conditional (DC) simulation method proposed in \cite{jovanovski2023towards}, extending it in two key directions: (i) to handle partially observed systems; (ii) to accommodate observations corrupted by measurement noise, in addition to the case of no measurement error. Before presenting the proposed DC method in detail, we briefly review the original approach and introduce the necessary notation. 

DC simulation \citep{jovanovski2023towards} improves the efficiency of ABC algorithms for parameter inference in SDE models, by guiding simulation paths toward the observed data. The method was originally developed for fully observed systems (one and two-dimensional) without measurement noise. For a given parameter value $\theta$, the approach begins by generating $P$ independent forward simulations of the SDE, denoted $\widetilde{X}^{1:P}_{1:d}$, over a fine grid $\tau_{0:nA}$. Each observational interval $ [t_i, t_{i+1}]$ is subdivided into $A$ equally spaced steps, so that $\tau_{iA + k} \in (t_i, t_{i+1}]$ for $k = 1, \ldots, A$. For each particle $j \in \{1, \ldots, P\}$, the simulated state of all $d$ chemical species at time $\tau_{iA + k}$ is denoted $\widetilde{X}^{j}_{1:d, iA + k}$. These intermediate states are assigned weights according to their proximity to the future observation $\Obs(t_{i+1})$, using a set of user-defined lookahead weighting functions $q(\Obs(t_{i+1}) \mid \widetilde{X}_{1:d, iA + k}^j)$. A backward-simulation particle smoother \citep{lindsten2013backward} is then applied to sample a single trajectory, resulting in a data-conditional path that closely mirrors the observations $\Obs$.

In this work, we modify the DC simulation scheme by removing the backward-simulation step and, for the case of particles corrupted with measurement noise, incorporating measurement noise directly into the weighting procedure. At each observation time $t_{i+1}$, we compute importance weights for the simulated states $\widetilde{Y}^j_{1:d_o, i+1}$ using a likelihood-based weighting function $q(\Obs(t_{i+1}) \mid \widetilde{Y}^j_{1:d_o, i+1})$, where $\widetilde{Y}^j_{1:d_o, i+1}$ denotes the simulated observation associated with particle $j$ (which may or may not include measurement noise). When the measurement noise is Gaussian, as assumed in this work, we take $q(\Obs(t_{i+1}) \mid \widetilde{Y}^j_{1:d_o, i+1}) = \phi_{d_{\rm o}}(\Obs(t_{i+1}) \mid \widetilde{Y}^j_{1:d_o, i+1}, C \Sigma^{\rm o})$, where $C$ is a scaling constant that expands the covariance. When the observables are not perturbed by measurement noise, a weighting function $q$ can be used, based on the EuM-induced transition density, see \cite{jovanovski2023towards} for details. It is important to note that this is different from what is done in the case of the bootstrap particle filter \citep{gordon1993novel}. Here, for $j \in \{1, ..., P\}$, the particles $\widetilde{Y}^j_{1:d_o, i}$ are assigned, at observational time $t_i$, a weight $\omega^j_{i}$ proportional to $q(\Obs(t_{i}) \given \widetilde{Y}^j_{1:d_o, i})$, instead of the latent particles $\widetilde{X}^j_{1:d, iA}$. Normalizing these weights yields a particle system $(\widetilde{Y}^{1:P}_{1:d_o, i}, \omega^{1:P}_{i})$ that approximates the densities
\begin{equation}
    \hat{p}_{\parm}({\rm d}y \given (\Obs(t_0), ..., \Obs(t_i)) = (\obs(t_0), ..., \obs(t_i))) = \sum_{j = 1}^P \omega_{i}^j \delta_{\widetilde{Y}^j_{1:d_o, {i}}} ({\rm d}y)
    \label{eq:part-approx-lookahead}
\end{equation}
where $\delta$ is the Dirac delta function. A single data-conditional trajectory is then constructed by sampling one particle per time point according to these weights, resulting in a path that reflects both the system dynamics and (if present) the noise in the observations. To generate a single data-conditional trajectory $Y^{\rm DC}$, we sample one particle at each observation time $t_i$ from the set $\widetilde Y^{1:P}_{1:d_o, i}$, using weights $\omega_i^{1:P} \propto q(\Obs(t_i) \mid \widetilde{Y}^j_{1:d_o, i})$. The selected value $\widetilde{Y}^j_i$ is then stored as $Y^{\rm DC}_{1:d_o, i} \coloneqq \widetilde{Y}^j_{1:d_o, i}$. For an algorithmic description of our novel method, see Algorithm~\ref{algo:forward-path}. In what follows, we define $Y^{\rm DC} \coloneqq Y^{\rm DC}_{1:d_o, 1:n}$ to lighten the notation.

\begin{algorithm}[ht]
\footnotesize
    \caption{Data-conditional path sampling $(\theta, C, h)$ for multivariate, partially observed SDEs}\label{algo:forward-path}
    \begin{algorithmic}[1]
        \FOR{$j = 1$ to $P$ in parallel} 
            \FOR{$i = 1$ to $n$}
                \FOR{$k = 1$ to $A$} 
                \STATE Simulate $\widetilde{X}_{1:d, (i-1)A + k}^j$ starting from $\widetilde{X}_{1:d, (i-1)A + k - 1}^j$ using the numerical integrator.
            \ENDFOR
            \IF{measurement noise}
            \STATE Sample $\widetilde{Y}^j_{1:d_o, i} \sim \mathcal{N}(\widetilde{X}^j_{1:d, iA}, \Sigma^{\rm o})$, and compute the weight $\omega_i^j \propto \phi_{d_{\rm o}}(\Obs(t_i) \given \widetilde{Y}^j_{1:d_o, i}, C\Sigma^{\rm o})$.
            \ELSE
            \STATE Set $\widetilde{Y}^j_{1:d_o, i} = L \widetilde{X}_{1:d, iA}$, and compute the weight $\omega^j_i$ based on the EuM-induced transition density (see \cite{jovanovski2023towards}).
            \ENDIF
            \ENDFOR
        \ENDFOR
        \FOR{$i = 1$ to $n$}
        \STATE At time $t_i$, normalize the particle weights $\omega^{1:P}_i$ and sample particle index $j \sim \text{Categorical}(\omega^{1:P}_i)$.
        \STATE Set $Y^{\rm DC}_{1:d_o, i} \coloneqq \widetilde{Y}^j_{1:d_o, i}$.
        \ENDFOR
    \STATE \textbf{Output:} Particle system $(\widetilde{Y}^{1:P}_{1:d_o, 1:n}, \omega^{1:P}_{1:n})$, and pseudo-observation $Y^{\rm DC}_{1:d_o, 1:n}$. 
    \end{algorithmic}
\end{algorithm}

\mycomment{
\begin{tikzpicture}
  \node[circle, fill, inner sep=2pt] (a1) at (0, 1) {};
  \node[circle, fill, inner sep=2pt] (a2) at (0,0.5) {};
  \node[circle, fill, inner sep=2pt] (a3) at (0,0) {};
    \node[circle, fill, inner sep=2pt] (a4) at (0, 1.5) {};
  \node[circle, fill, inner sep=2pt] (a5) at (0,2) {};

    \node[circle, fill, inner sep=2pt] (b1) at (1, 1) {};
  \node[circle, fill, inner sep=2pt] (b2) at (1,0.5) {};
  \node[circle, fill, inner sep=2pt] (b3) at (1,0) {};
    \node[circle, fill, inner sep=2pt] (b4) at (1, 1.5) {};
  \node[circle, fill, inner sep=2pt] (b5) at (1,2) {};

    \node[circle, fill, inner sep=2pt] (c1) at (2, 1) {};
  \node[circle, fill, inner sep=2pt] (c2) at (2,0.5) {};
  \node[circle, fill, inner sep=2pt] (c3) at (2,0) {};
    \node[circle, fill, inner sep=2pt] (c4) at (2, 1.5) {};
  \node[circle, fill, inner sep=2pt] (c5) at (2,2) {};

    \node[circle, fill, inner sep=2pt] (d1) at (3, 1) {};
  \node[circle, fill, inner sep=2pt] (d2) at (3,0.5) {};
  \node[circle, fill, inner sep=2pt] (d3) at (3,0) {};
    \node[circle, fill, inner sep=2pt] (d4) at (3, 1.5) {};
  \node[circle, fill, inner sep=2pt] (d5) at (3,2) {};

      \node[circle, fill, inner sep=2pt] (e1) at (4, 1) {};
  \node[circle, fill, inner sep=2pt] (e2) at (4,0.5) {};
  \node[circle, fill, inner sep=2pt] (e3) at (4,0) {};
    \node[circle, fill, inner sep=2pt] (e4) at (4, 1.5) {};
  \node[circle, fill, inner sep=2pt] (e5) at (4,2) {};
  
  \node[circle, fill, inner sep=2pt] (b) at (1,1) {};
  \node[circle, fill, inner sep=2pt] (c) at (2,0.5) {};
  \node[circle, fill, inner sep=2pt] (d) at (3,1.5) {};

  \draw (a1) -- (b1);
  \draw (a2) -- (b2);
  \draw (a3) -- (b3);
  \draw (a4) -- (b4);
  \draw (a5) -- (b5);
  
  \node[circle, draw, inner sep=3pt] (x) at (2,2) {};
  \draw (x) node[cross out, draw=black] {};

\end{tikzpicture}}

In \cite{jovanovski2023towards}, the integration of the DC simulator into ABC-SMC produced the ABC-SMC-DC Algorithm \ref{algo:sl-abs-smc} (in Supplementary Material), which has led to substantial improvements in both speed and accuracy of the posterior approximation. Much of this improvement is due to (a) the close alignment between the simulated path and the observed data, particularly for parameters drawn from regions of high-posterior density, and (b) an  importance sampling correction. In this work, we have further improved ABC-SMC-DC by replacing the EuM scheme with our new splitting scheme,
accelerating inference for challenging models where the simulated paths are highly erratic and oscillating, as for the Repressilator and Lotka-Volterra models. This is enabled by the structure preserving properties of our splitting scheme.
 Given the above, recall that in ABC the simulations from the data-generating model are typically compressed into summary statistics. Now, the summary statistic obtained by simulating a data-conditional trajectory and summarizing it, is denoted with $S^{\rm DC}\equiv S(Y^{\rm DC})$, and we write $S^{\rm DC} \sim p_\theta(S^{\rm DC} \given Y^o)$. The latter notation may appear slightly redundant, as a data-conditional quantity is by definition conditional to $Y^o$, however we employ this notation to help the distinction with $p_\theta(S^{\rm DC})$, which we use to denote the density of the summaries produced by the \textit{forward} simulator and evaluated in $S^{\rm DC}$. In fact,  since the distribution of the DC-summary $p_\theta(S^{\rm DC} \given Y^o)$ is different from that of the standard forward simulator $p_\theta(S)$, the weight of an accepted parameter particle needs to be adjusted by the importance ratio $\ptran(S^{\rm DC}) / \ptran(S^{\rm DC} \given \Obs)$. While this importance ratio is intractable, it can be efficiently approximated by replacing the two densities with  their corresponding ``synthetic likelihood''  approximations \citep{wood2010statistical}. Specifically, we approximate $p_\theta(S) \approx \mathcal{N}(S \given \mu_P, \Sigma_P)$ and $p_\theta(S^{\rm DC} \given \Obs) \approx \mathcal{N}(S \given {\mu}^{\rm DC}_{P, \parm}, {\Sigma}^{\rm DC}_{P, \parm})$, where $\mu_{P, \parm}, \Sigma_{P, \parm}$ are the empirical mean and covariance matrix of $P$ summaries simulated independently as $S^{1:P} \sim p_\theta(S)$, and similarly ${\mu}^{\rm DC}_{P, \parm}, {\Sigma}^{\rm DC}_{P, \parm}$ are the empirical mean and covariance matrix of ${S}^{\textrm{DC}, 1:P} \sim p_\theta(S^{\rm DC} \given \Obs)$. Implicitly, the distribution of the data-conditional path depends on the particle system $(Y^{1:P}_{1:d_o, 1:n}, w^{1:P}_{1:n})$ from the complete forward paths (the weighted system is necessary to build the DC path), 
 i.e. $p_\theta({Y}^{\rm DC} \given \Obs) \approx \hat{p_\theta}({Y}^{\rm DC} \given \Obs, (Y^{1:P}_{1:d_o, 1:n}, w^{1:P}_{1:n}))$. Therefore, the samples ${S}^{\textrm{DC}, 1:P}$ are obtained by sampling paths from the joint filtering density and summarizing them. Now, recall that in our work the summaries are expressed via sequentially trained neural networks: then, at round $r$ of ABC-SMC-DC, for a sequentially retrained summary $S_r(\cdot)$, the approximate parameter particle weight is given by 
\begin{equation}
        w_r({\theta}, S_r({{Y}^{\rm DC}})) \appropto \frac{\mathbbm{1}( \Vert S_r({{Y}}^{\rm DC}) - {S}_r(Y^{\rm o}) \Vert < \epsilon_r)\pi({\theta})}{\sum_{i=1}^M W^i_{r - 1} \phi_p({\theta} \given {\theta}_{r - 1}^i, {\Sigma}_{r - 1})} \frac{\phi_p(S_r({{Y}}^{\rm DC}) \given {\mu}_{P, {\theta}}, {\Sigma}_{P, {\theta}}) }{\phi_p(S_r({{Y}}^{\rm DC}) \given {{\mu}}^{\rm DC}_{P, {\theta}}, {{\Sigma}}^{\rm DC}_{P, {\theta}})} \label{eq:seq-imp-weight}.
\end{equation}
The (unnormalized) weights \eqref{eq:seq-imp-weight} are computed for a set of the $M$ parameter-summary pairs, replacing the weights \eqref{eq:particle-weight-abcsmc} used in standard (forward-only) ABC-SMC. Notably, \eqref{eq:seq-imp-weight} reduces to \eqref{eq:particle-weight-abcsmc} in the case of forward-only simulations, as no synthetic likelihood correction is needed there. The indicator function $\mathbbm{1}( \Vert S_r({Y}^{\rm DC}) - {S}_r(Y^{\rm o} \Vert < \epsilon_r) $ denotes that weights are computed only if the pair $(\theta, S_r({Y}^{\rm DC})) $ is accepted, a condition that is implicitly handled in \eqref{eq:particle-weight-abcsmc}.\\

\section{Simulation studies}
\label{sec:simulation-studies}

\textcolor{black}{In the following, we evaluate the performance of our proposed numerical splitting scheme and ABC-SMC-DC inference algorithm on three widely used chemical reaction network models, namely the stochastic Repressilator, the stochastic Lotka--Volterra model, and the Two-pool model. First, we investigate whether some important structural properties of these models, such as their oscillatory behavior, mean dynamics and asymptotic distributions, are preserved by our splitting schemes and by the EuM method \eqref{eq:EuM} for increasing time steps $h$. Then, we estimate parameters of interest in the case of partial and/or noisy observations, comparing our ABC-SMC-DC algorithm with the standard \lq\lq forward-only\rq\rq ABC-SMC, which simulates SDE paths without data-conditioning. Both algorithms run for up to 20 iterations, unless the acceptance rate (defined as the ratio between the number of accepted and sampled particles) falls below 1.5\%,  in which case they are immediately terminated.} In all experiments, summary statistics are outputted by the partially exchangeable network (PEN), which is pre-trained on 50,000 prior-predictive samples. To accommodate the retraining of PEN, the number of parameter particles is set to 10,000.

\subsection{A genetic oscillator: stochastic Repressilator}

\label{sec:stoch-Repressilator}

We investigate the dynamics of the Repressilator model \citep{elowitz2000synthetic}, a synthetic genetic regulatory network comprising three genes, each inhibiting the next to form a cyclic negative feedback loop. This configuration gives rise to oscillatory behavior. The system is modeled through reactions that govern the transcription of three mRNAs ($M_1$, $M_2$, $M_3$), their translation into proteins ($P_1$, $P_2$, $P_3$), and their subsequent degradation. The inhibition mechanism of each gene involves nonlinear repression modulated by a Hill coefficient. For the stochastic version of the Repressilator, we follow the exposition in \cite{warne2020practical}. This is a 6-dimensional SDE comprising two sets of similar SDEs: one set for mRNA and another for protein levels. Within each set, the three SDEs exhibit structural similarity but differ in specific parameters. The interactions of proteins $P_i$ within the mRNA SDEs do not follow mass-action kinetics; however, \textcolor{black}{conditionally on the proteins being fixed}, 
the equations for $M_i$ are of CIR-type. Similarly, the SDEs governing the protein levels also follow CIR-type dynamics, when $M_i$ are held fixed. The SDE system for $i = 1, 2, 3$ is as follows:
\begin{eqnarray}\nonumber 
    \rd M_i(t) &=& \left( \alpha_0 + H(P_j(t)) - M_i(t) \right) \rd t + \sqrt{\alpha_0 + H(P_j(t))} \, \rd W^{(4i - 3)}(t) - \sqrt{M_i(t)} \, \rd W^{(4i)}(t),   \\ 
    &&\label{repr}\\ \nonumber
    \rd P_i(t) &=& \beta \left( M_i(t) - P_i(t) \right) \rd t + \sqrt{\beta M_i(t)} \, \rd W^{(4i - 2)}(t) - \sqrt{\beta P_i(t)} \, \rd W^{(4i - 1)}(t), \  
\end{eqnarray}
where $j = (i + 1) \mod 3 + 1$,  $H(P_j) = \alpha K ^ n / (K^n + P_j^n)$ is the Hill function, $\alpha, \gamma, \alpha_0,\beta,K\in\mathbb{R}_+$. We refer to Section \ref{SectionC1} in the Supplementary Material, for a detailed derivation of a Strang splitting scheme for this model, following the procedure introduced in Section \ref{sec:cond-cir}.

\subsubsection{Preservation of distribution}
\begin{figure}[ht]
    \centering
    \includegraphics[width=0.5\linewidth]{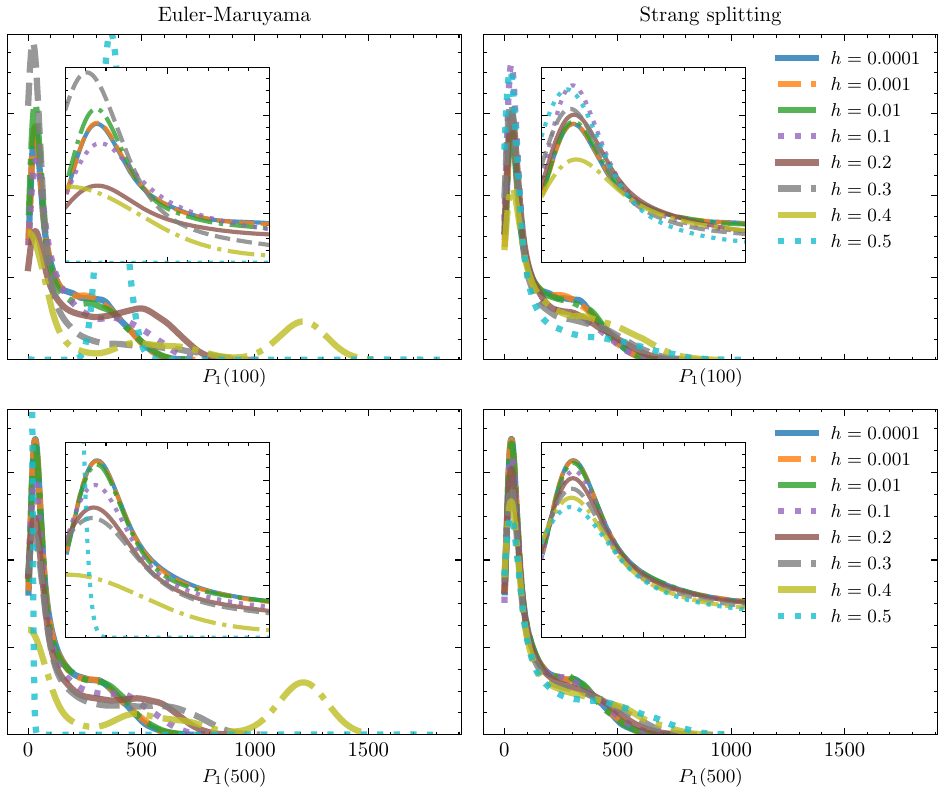}
    \caption{
    Stochastic Repressilator model \eqref{repr}. Kernel density estimates of the distribution of $P_1(t)$ at time $t=100$ (top panels) and $t=500$ (bottom panels), when simulating trajectories with EuM (left panels) and Strang (right panels) for different time steps.}  

    \label{fig:Repressilator-dist-preserve}
\end{figure}

Here, we investigate the ability of EuM and of the derived Strang splitting scheme to preserve the distribution of the process at time $t=100$ and $t=500$ (
the latter in the asymptotic/invariant/stationary regime), for different time steps $h$, ranging from $10^{-4}$ to $0.5$. To do this, 20,000 paths were simulated for each integration time step and scheme, recording values at $t=100, 500$, which were then used to compute the kernel density estimator (KDE) of the $P_1$ component, reported in Figure \ref{fig:Repressilator-dist-preserve}. While the distributions generated by the EuM method show considerable variation as the time step increases, those generated by Strang remain relatively stable, indicating a more consistent preservation of the distribution.

\subsubsection{Inference with splitting versus EuM schemes}
\begin{figure}
    \centering
    \includegraphics[height=0.6\linewidth]{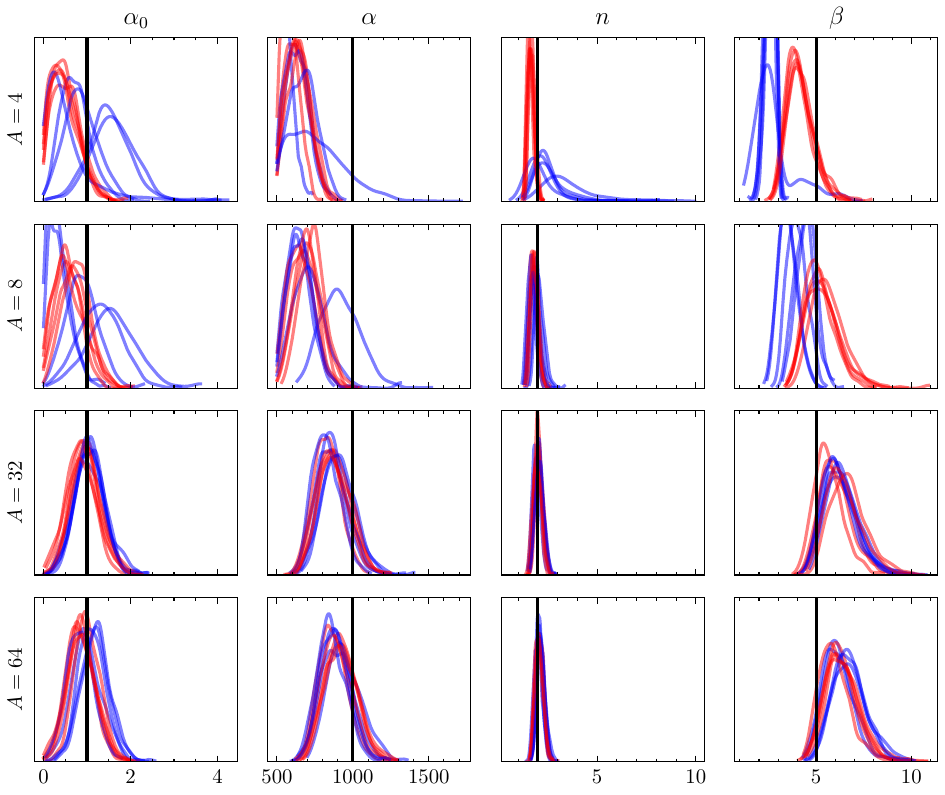}
    \caption{
    Stochastic Repressilator model \rc{\eqref{repr}}: marginal posterior distributions using EuM (blue) vs Splitting (red). Rows represent the number of inter-observational intervals and columns represent parameters.}
    \label{fig:em-vs-strang-Repressilator}
\end{figure}

To evaluate the performance of the proposed splitting integrator in comparison with the classical EuM method, we consider a simulation-based inference scenario using ABC-SMC on a fully observed system. The true latent state is observed at $n = 50$ equally spaced time points over the interval $[0, 100]$, resulting in an observation interval $\Delta = 2$. The synthetic data is generated with the true parameter values, and inference is performed using marginal ABC-SMC with uniform prior components: $\alpha_0 \sim \mathcal{U}(0, 10)$, $\alpha \sim \mathcal{U}(500, 2500)$, $n \sim \mathcal{U}(0, 10)$, and $\beta \sim \mathcal{U}(0, 20)$. To simulate the forward trajectories, we fix the coarse observation grid and vary the number of integration subintervals between observations as $A \in \{4, 8, 32, 64\}$, corresponding to integration steps $h = \Delta  / A$. As shown in Figure \ref{fig:em-vs-strang-Repressilator},
the proposed splitting scheme yields stable posterior inference even at coarse integration steps (e.g., $A = 4$, $h = 0.5$), whereas EuM results in biased and unstable posteriors. For larger values of $A$, both integrators stabilize; however, the splitting scheme consistently produces sharper and more accurate posteriors across all tested values. 
\subsubsection{Inference of kinetic parameters and hill coefficients from partial observations}
\begin{figure}[ht]
    \centering
    \includegraphics[width=0.7\linewidth]{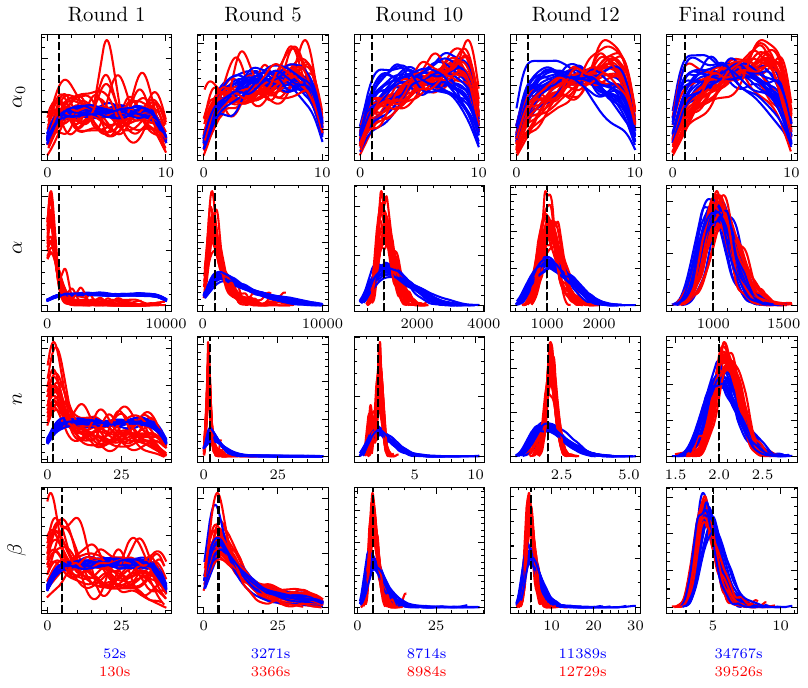}
    \caption{Stochastic Repressilator model \eqref{repr}. Marginal posterior distributions at several rounds of ABC-SMC (blue lines) and ABC-SMC-DC (red lines), for 20 repeated runs. 
    The true parameter values are displayed as black vertical lines. Below each panel, the cumulative runtimes (in seconds) of the algorithms, averaged over 20 runs, are displayed.}
    \label{fig:Repressilator-inference}
\end{figure}

Among the six components, 
we assume that only the proteins are observed, not directly, but with measurement error. That is, the observational model is  given by
$$
(Y^{\rm o}_1, Y^{\rm o}_2, Y^{\rm o}_3) = (P_{1, i}, P_{2, i}, P_{3, i}) + \xi_i, \quad \xi_i \sim \mathcal{N}_3(0, \sigma^2_{\text{err}} I),
$$  
where $P_{1, i}$, $P_{2, i}$, and $P_{3, i}$ represent the observed protein concentrations, and $\xi_i$ is Gaussian noise with diagonal covariance matrix $\sigma^2_{\text{err}} I$, where $\sigma_{\text{err}}$ is assumed known. We generate a discrete realization of length 50 on the interval $[0, 10]$, resulting in a sampling rate of $\Delta = 0.2$. The system starts at time $t_0 = 0$ with known initial state $X_0 = (0, 40, 0, 20, 0, 60)$ parameter $K=20$ and measurement standard deviation error $\sigma_{\text{err}} = 5$, while the unknown parameters $\theta=(\alpha_0,\alpha,n,\beta)$, which we aim to estimate, are set to $(1,1000,2,5)$. The number of subintervals $A$ is set to 10, therefore $h = 0.02$, and the scaling constant for the covariance in the data-conditional simulator is set to $C = 20$. 
We adopt an independent prior specification and set  
$$
\pi(\theta) = \mathcal{U}(\alpha_0 \mid 0, 10) \mathcal{U}(\alpha \mid 0, 10^4) \mathcal{U}(n \mid 0, 40) \mathcal{U}(\beta \mid 0, 40),
$$  
where $\mathcal{U}(\cdot|a,b)$ denotes the pdf of a continuous uniform distribution in $(a,b)$.
The marginal posterior approximations across different iterations/rounds for both standard ABC-SMC (blue lines) and ABC-SMC-DC (red lines) are shown in Figure \ref{fig:Repressilator-inference}.  
Notably, even in the first round of ABC-SMC-DC, the posteriors for $\alpha$ and $n$ are already peaked around the true values, indicating the algorithm’s ability to guide parameter towards regions of high posterior density, which is not the case for standard ABC-SMC. By round 5,  a similar trend is observed for $\beta$. In contrast, ABC-SMC exhibits slower convergence, taking more rounds to achieve comparable results. By round 10, the marginal ABC-SMC-DC posteriors 
begin to stabilize, suggesting that further iterations provide diminishing returns. At this stage, the algorithm can be reasonably stopped. Importantly, while $\alpha_0$ remains unidentifiable with both algorithms, the results indicate that $\alpha$, $n$ and $\beta$ can be inferred effectively. If run long enough, both algorithms ultimately converge to similar posteriors for all parameters, with the ABC-SMC characterized by a higher runtime than ABC-SMC-DC, and recall that for ABC-SMC-DC we could have safely stopped at round 10 with a much smaller runtime, resulting in a 4-folds acceleration.

\subsection{Predator-prey model: stochastic Lotka--Volterra}
The Lotka-Volterra model describes the interaction between predator and prey populations. Denote by \(\X(t) = (X_1(t), X_2(t))\) the populations of predators \(X_1(t)\) and prey \(X_2(t)\) at any given time \(t\). The Stochastic Lotka--Volterra model (e.g. \citealp{golightly2011bayesian}) is defined by
\begin{subequations}
    \begin{align}
    \rd X_1(t) & = (\theta_1 - \theta_2 X_2(t)) X_1(t) \, \rd t + \sqrt{\theta_1 X_1(t)} \, \rd W^{(1)}(t) - \sqrt{\theta_2 X_1(t) X_2(t)} \, \rd W^{(2)}(t), \label{eq:lv-sde1}\\
    \rd X_2(t) & = (\theta_2 X_1(t) - \theta_3) X_2(t) \, \rd t + \sqrt{\theta_2 X_1(t) X_2(t)} \, \rd W^{(2)}(t) -\sqrt{\theta_3 X_2(t)} \, \rd W^{(3)}(t). \label{eq:lv-sde2}
\end{align}
\end{subequations}
Unlike for the Repressilator model, the species share a Brownian motion, $W^{(2)}(t)$, so the Brownian motions cannot be combined into a unique Gaussian term (therefore we are in Scenario 1 in Section \ref{sec:subeq}). To apply our splitting scheme to this system, we should first rewrite SDE  \eqref{eq:lv-sde1} for species $X_1$ as \eqref{eq:cle-cond-gen}. We refer to Section \ref{SectionC2} of the Supplementary Material for the derivation of a Lie-Trotter scheme for this model.

\subsubsection{Preservation of oscillatory behaviour} 

\mycomment{\begin{equation}
X_{1, k + 1} = \left(e^{-\widetilde{b}_1 h / 2} \sqrt{X_{1, k} - \frac{\widetilde{c}_{1, 1}^2 + \widetilde{c}_{1, 2}^2}{4}h} + \sqrt{\frac{1 - e^{-\widetilde{b}_1 h}}{4 \widetilde{b}_1}} \left(c_{1, 1} \Delta W^{(1)}_{k+1} + c_{1, 2} \Delta W^{(2)}_{k+1}\right)\right)^2. \label{eq:lv-cond-split1}
\end{equation}}

\begin{figure}
    \centering
    \includegraphics[scale=1]{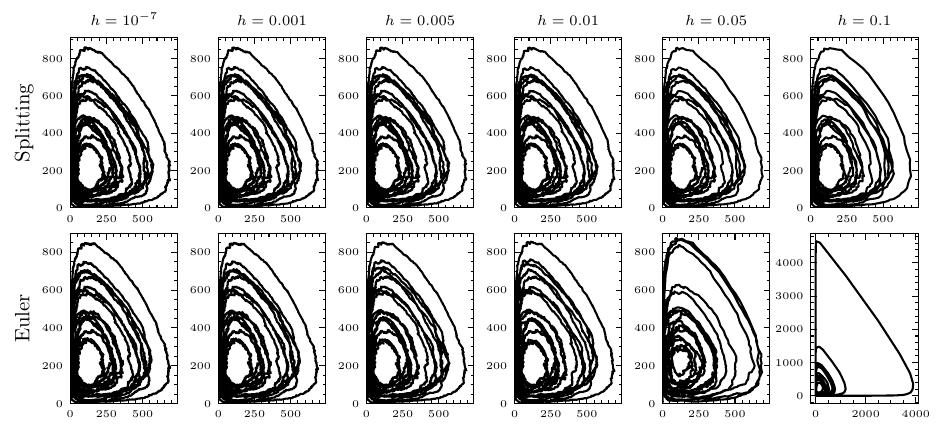}
    \caption{
    Stochastic Lotka-Volterra model \eqref{eq:lv-sde1}-\eqref{eq:lv-sde2}: Comparison of phase portraits obtained using the derived splitting scheme (top panels) and EuM (bottom panels) for increasing integration time steps $h$. The leftmost column presents the solution of both schemes with a time step of \(10^{-7}\). As the integration time step reaches \(h = 0.05\), a noticeable difference between the two schemes emerges, and eventually EuM breaks down at \(h = 0.1\).}
    \label{fig:lv-oscillations}
\end{figure}

\mycomment{\eqref{eq:lv-sde2} is given by:
\begin{equation}
X_{2, k + 1} = \left(e^{-\widetilde{b}_2 h / 2} \sqrt{X_{2, k} - \frac{\widetilde{c}_{2, 2}^2 + \widetilde{c}_{2, 3}^2}{4}h} + \sqrt{\frac{1 - e^{-\widetilde{b}_2 h}}{4 \widetilde{b}_2}} \left(\widetilde{c}_{2, 2} \Delta W^{(2)}_{k+1} + \widetilde{c}_{2, 3} \Delta W^{(3)}_{k+1}\right)\right)^2. \label{eq:lv-cond-split2}
\end{equation}}
Figure \ref{fig:lv-oscillations} highlights the comparative performance of the splitting and EuM schemes when looking at the phase portrait of the model for $\theta=(\theta_1,\theta_2,\theta_3)=(0.5,0.0025,0.3)$, since the model oscillates for this configuration. Our scheme consistently maintains accuracy whereas EuM degrades in accuracy at 0.005, and even fails at 0.1.
\mycomment{\textcolor{red}{Taylor 1.5 for the synaptic conductance model vs our splitting scheme. We need to show that oscillations are not preserved and then we are golden. Maybe we can also do it for the invariant distribution of the harmonic oscillator? Show that it may be not preserved with the Taylor 1.5 scheme, but it will be with our scheme.}}

\subsubsection{Inference of kinetic parameters from complete observations}

For the experimental scenario, we assume that all components of the Lotka-Volterra system are observed with measurement error. In particular, the observational model is given by
$$
\Obs_i = X_i + \xi_i, \quad \xi_i \sim \mathcal{N}_2(0, \sigma^2_{\text{err}} I),
$$  
where $X_i = (X_{1, i}, X_{2, i})$ represents the true system states at time $t_i$. Here $\sigma_{\text{err}}$ is assumed known and equal to $\sigma_{\text{err}} = 10$. The system is initialized with the state $X_0 = (100, 100)$, and the true parameter values are $\theta = (0.5, 0.0025, 0.3)$. A discrete realization of length 50 is generated over the interval $[0, 50]$, giving a sampling rate of $\Delta = 1$. The number of subintervals $A$ is set to $100$, giving an integration timestep $h = 0.01$, and the scaling constant for the covariance in the data-conditional simulator is set to $C = 20$. We take an independent prior specification, with component distributions chosen to cover plausible ranges, as in \cite{golightly2011bayesian}:  
$$
\pi(\theta) = \mathcal{U}(\theta_1 \mid 0, 1) \mathcal{U}(\theta_2 \mid 0, 0.05) \mathcal{U}(\theta_3 \mid 0, 1).
$$  

\begin{figure}[ht]
    \centering
    \includegraphics[width=0.7\linewidth]{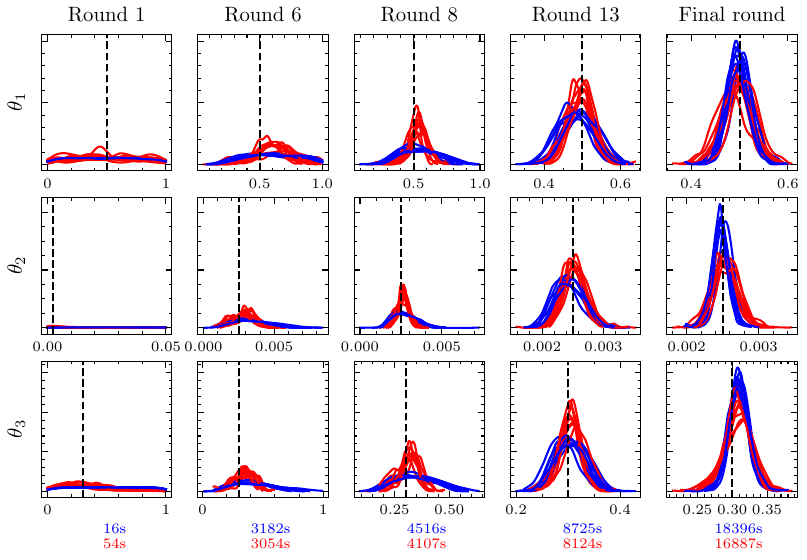}
    \caption{Stochastic Lotka--Volterra model \rc{\eqref{eq:lv-sde1}-\eqref{eq:lv-sde2}}: Marginal posterior distributions at several rounds of ABC-SMC (blue lines) and ABC-SMC-DC (red lines). 
    The true parameter values are displayed as black vertical lines. Under each panel, the cumulative runtimes (in seconds) of the algorithms, averaged over 20 runs, are displayed}. 
    \label{fig:lv-inference}
\end{figure}

The marginal parameter posteriors from ABC-SMC and ABC-SMC-DC
are shown in Figure~\ref{fig:lv-inference}.
By round 6, the posterior distributions inferred using ABC-SMC-DC begin to concentrate in regions of high posterior density, whereas those obtained with ABC-SMC shows slower progress towards these regions. By round 8, the posterior distributions for ABC-SMC-DC appear to concentrate around the true parameter values, whereas those from ABC-SMC are just beginning to shift towards them.  
At round 13, the posterior distributions produced by ABC-SMC finally becomes comparable to those produced by ABC-SMC-DC at round 8, implying that ABC-SMC-DC achieves these results in nearly half the time.  

\subsection{Two-pool model}
The two-pool model describes the decay of a substance that is able to transfer between two pools \citep{nestel1969distribution}. The model can be described as a chemical reaction network by
\begin{equation}
    \mathcal{R}_1: X_1 \xrightarrow{\theta_1} \emptyset, \quad \mathcal{R}_2: X_2 \xrightarrow{\theta_2} \emptyset, \quad \mathcal{R}_3: X_1 \xrightarrow{\theta_3} X_2, \quad \mathcal{R}_4: X_2 \xrightarrow{\theta_4} X_1, \nonumber 
\end{equation}
with stoichiometries $\nu_1 = (-1, 0)$, $\nu_2 = (0, -1)$, $\nu_3 = (-1, 1)$, $\nu_4 = (1, -1)$, and propensities $a_1(X(t)) = \theta_1 X_1(t)$, $a_2(X(t)) = \theta_2 X_2(t)$, $a_3(X(t)) = \theta_3 X_1(t)$, $a_4(X(t)) = \theta_4 X_2(t)$. The CLE is given by:
\begin{subequations}
\begin{align}
    \rd X_1(t) & = \left(\theta_4 X_2(t) - (\theta_1 + \theta_3) X_1(t)\right) \rd t \label{eq:twopool-1}\\
               & \quad -\sqrt{\theta_1 X_1(t)} \, \rd W^{(1)}(t) -\sqrt{\theta_3 X_1(t)} \, \rd W^{(3)}(t) + \sqrt{\theta_4 X_2(t)} \, \rd W^{(4)}(t), \nonumber \\
    \rd X_2(t) & = \left(\theta_3 X_1(t) - (\theta_2 + \theta_4) X_2(t)\right) \rd t \label{eq:twopool-2}\\
               & \quad -\sqrt{\theta_2 X_2(t)} \, \rd W^{(2)}(t) + \sqrt{\theta_3 X_1(t)} \, \rd W^{(3)}(t) - \sqrt{\theta_4 X_2(t)} \, \rd W^{(4)}(t). \nonumber 
\end{align}
\end{subequations}
The Brownian motion cannot be combined into a unique Gaussian term (Scenario 1 in Section \ref{sec:subeq}), because the species share two Brownian motions, $W^{(3)}$ and $W^{(4)}$. To apply our splitting scheme to this system, we transform the SDE for species $X_1$ \eqref{eq:twopool-1} into the form specified in \eqref{eq:cle-cond-gen}. We refer to Section \ref{SectionC3} of the Supplementary Material for the derivation of a Lie-Trotter scheme for this model.

\subsubsection{Inference of kinetic parameters and measurement noise from partial observations}
\begin{figure}[ht]
    \centering
    \includegraphics[width=0.7\linewidth]{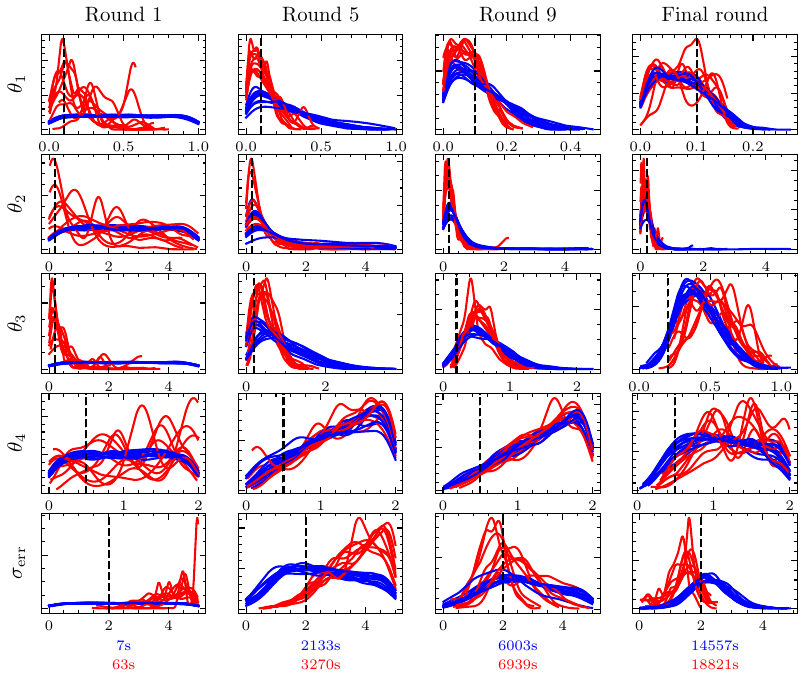}
    \caption{Two-pool model \eqref{eq:twopool-1}-\eqref{eq:twopool-2}. Marginal posterior distributions at several rounds of ABC-SMC (blue lines) and ABC-SMC-DC (red lines). The true parameter values are displayed as black vertical lines. Under each panel, the cumulative runtimes (in seconds) of the algorithms, averaged over 10 runs, are displayed.}
    \label{fig:twopool-inference}
\end{figure}

We consider an experimental scenario similar to that of \cite{browning2020identifiability}, where the available observations $\Obs_i$ consist of discrete-time measurements of the first components with additional measurement noise, that is $\Obs_i = X_{1, i} + \xi_i$, with $\xi_i \sim \mathcal{N}(0, \sigma_{\text{err}}^2)$.
The process starts at $X_0 = (100, 0)$, and the observations are generated using the parameters 
$\theta=(\theta_1,\theta_2,\theta_3,\theta_4,\sigma_{\text{err}})=(0.1, 0.2, 0.2, 0.5,2)$. Unlike in previous examples, here the standard deviation $\sigma_{\text{err}}$ is assumed to be unknown. 
The observed trajectory consists of 50 equidistant data points, collected with a time step of $\Delta = 0.2$. The number of subintervals $A$ is set to $10$, yielding $h 
= 0.02$, and the scaling constant for the covariance in the data-conditional simulator is set to $C = 2$. The parameter prior is defined as:  
$$
\pi(\theta) = \mathcal{U}(\theta_1 \mid 0, 1) \mathcal{U}(\theta_2 \mid 0, 5) \mathcal{U}(\theta_3 \mid 0, 5) \mathcal{U}(\theta_4 \mid 0, 2) \mathcal{U}(\sigma_{\text{err}} \mid 0, 5).
$$
We chose slightly more diffuse prior components than \cite{browning2020identifiability} to challenge ABC-SMC and better display the acceleration towards the posterior brought by ABC-SMC-DC. The posterior approximations for both ABC-SMC and ABC-SMC-DC are shown in Figure \ref{fig:twopool-inference}. At convergence they are close to the posteriors in Figure 6 of \cite{browning2020identifiability}, which use MCMC. 
Already at round 1, the ABC-SMC-DC algorithm shows early progress, with the posterior distributions for $\theta_1$ and $\theta_3$ starting to concentrate near the true parameter values. By round 5, the algorithm is sampling from regions of high posterior density for $\theta_1$, $\theta_2$, and $\theta_3$, although not for $\sigma_{\text{err}}$. By round 9, the posterior distributions for all parameters, except $\sigma_{\text{err}}$, appear to have stabilized. Further iterations yield diminishing returns, suggesting that the ABC-SMC-DC algorithm can reasonably be stopped at this stage.  

\section{Discussion}\label{Section7}
We contribute to two important problems in applied mathematics, i.e., (i) the construction of reliable and efficient numerical schemes for multivariate SDEs, and (ii) the inference of model parameters from partially and noisily observed SDEs. To achieve this, we first derived what we called (perturbed) conditionally-CIR-type models, a large class of SDEs  where each component, conditionally on the others being fixed, behaves like a perturbed CIR (square-root) process. This is a very broad model class, which we show to include several well known biochemical reaction network models,  which can be formulated through CLEs, such as the Repressilator, Lotka-Volterrra and the two-pool models.

On the numerical side (i), we derive an innovative numerical scheme, embedding the Lamperti transform and conditional approaches into splitting schemes, further developing and generalizing what has been recently proposed for univariate CIR processes \citep{kelly2023adaptive} to the more challenging setting of multivariate SDEs with several shared joint square-root diffusion components. 
Unlike the commonly used Euler-Maruyama (EuM) method, our scheme succeeds in preserving important structural properties of the models, such as oscillatory dynamics, state-space (e.g. non-negativity)  and invariant distributions, even for large time steps. This is a crucial feature when embedding a numerical scheme in simulation-based inference, allowing to reduce the computational cost and runtime of the algorithms.

On the inferential side (ii), we propose a variant of the ABC-SMC-DC initially presented in \cite{jovanovski2023towards} for fully observed SDEs and no measurement noise, with three major updates. First, the new algorithm accommodates partially observed as well as noisily observed SDEs. Second, it relies on summary statistics derived from a partially exchangeable network (PEN) based on non-Markovian observations, extending the PEN introduced in \cite{wiqvist2019partially} for Markov processes. 
Third, it is based on the derived splitting scheme instead of EuM. As a result, our data-conditional ABC inference algorithm, unlike most simulation-based inference methods, does not simulate a realization from the model in a ``myopic'' way, but rather uses the observed data to produce model solutions that are ``data-informed''. 
This result in a computationally efficient franework, compared to the canonical (forward-only) ABC-SMC, giving a reduction in computational cost of around a factor of 2-4, depending on the application.

Although we have chosen to focus on CRNs, we emphasise that the proposed numerical and inferential methods can be applied to any SDE which can be rewritten as a perturbed conditionally-CIR-type model. While broad, this class of SDEs does not include those CLEs which are conditionally non-linear in the drift, such as the Schl{\"o}gl model. This would require a different numerical scheme, whose derivation and impact when embedded within simulation-based inference merits the attention of future work.

\section*{Acknowledgments}
UP acknowledges funding from the Swedish National Research Council (Vetenskapsrådet 2019-03924). PJ and UP acknowledge funding from the Chalmers AI Research Centre.

\bibliographystyle{abbrvnat}  
\bibliography{main}

\clearpage
\appendix

\begin{center}
    \LARGE
    Supplementary Material
\end{center}

\section{Numerical splitting schemes for conditionally linear ODEs}\label{CLODE}
Consider a system of ODEs
\begin{equation}\label{ODE}
\rd x(t)=f(x(t))\rd t,
\end{equation}
where $f: \mathbb{R}^d \to \mathbb{R}^d$ is the vector field with components $f_i(x(t)) = a_i(x)x_i(t) + b_i(x(t))$, for $i = 1, \ldots, d$, i.e.
\begin{equation}
    \rd x_i(t) = (a_i(x(t)) x_i(t) + b_i(x(t))) \, \rd t, \quad i = 1, \ldots, d, \label{eq:cls-system}
\end{equation}
where $a_i$, $b_i$ are functions depending on $x_j$ for $j \neq i$. These systems have the property that, if all $x_j, j\neq i$ are fixed (i.e., they are considered to be constant), then $x_i$ satisfies a first-order linear ODE, which can be solved exactly. Hence,  \cite{chen2020structure} proposed to split \eqref{ODE} as
\begin{equation}\label{eq}
\rd x^{(j)}(t)=f^{(j)}(x(t)) \, \rd t, \quad j=1,\ldots, d,
\end{equation} 
where the sub-vector fields $f^{(j)}: \mathbb{R}^d \to \mathbb{R}^d$ are given by 
\begin{equation}
    f^{(j)}_i(x(t)) = 
    \begin{cases}
        f_i(x{(t)}){=a_i(x{(t)})x_i{(t)}+b_i(x{(t)})} & \text{if } i = j, \\
        0 & \text{if } i \neq j,
    \end{cases}
    \label{eq:chen-splitting}
\end{equation}
such that $f = f^{(1)} + \ldots + f^{(d)}$. Hence, when solving \eqref{eq}, all components of $x_i^{(j)}, i\neq j$ are constant, while that for $i=j$ is obtained by solving the linear ODE in  \eqref{eq:chen-splitting}. Subsequently, the solution of \eqref{ODE} in $\tau_k$ starting at time $\tau_{k-1}$ can be obtained by composing the flows (solutions) of the $d$ subequations \eqref{eq} via the Lie-Trotter or Strang compositions \citep{mclachlan2002splitting}. The obtained splitting scheme effectively preserves limit cycles, see \cite{chen2020structure}.

\subsection{Illustration for (deterministic) CRNs}
\begin{figure}[ht]
    \centering
    \includegraphics[width=0.9\linewidth]{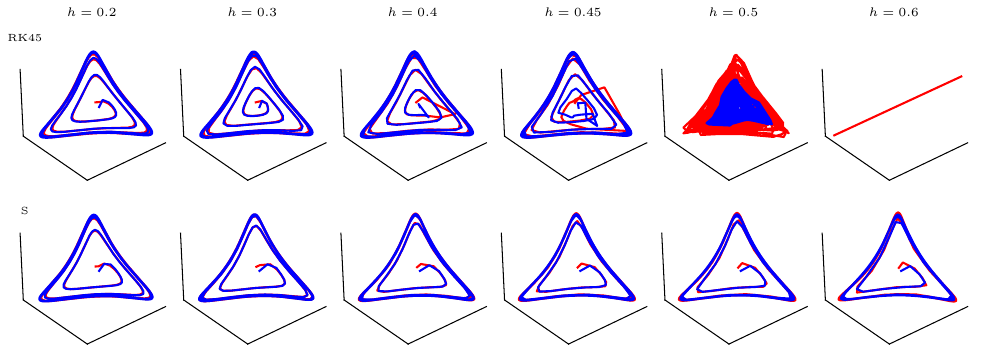}
    \caption{Deterministic Repressilator model: 
    numerical solution of the ODEs \eqref{eqP}-\eqref{eqM} with different time steps $h$ using the Runge-Kutta method (top row) and 
    conditionally linear splitting with Strang composition (bottom row). The 3D trajectories of mRNAs $(M_1(t), M_2(t), M_3(t))$ are shown in blue, and the 3D trajectories of proteins $(P_1(t), P_2(t), P_3(t))$ are shown in red. It can be observed that the  limit cycle is preserved by the splitting scheme even at large time steps.}
    \label{fig:rk-vs-split-limitcycle}
\end{figure}

CRNs can also be described using ODEs, which provide a macroscopic, deterministic view of the dynamics. Interestingly, many ODE descriptions of CRNs exhibit a conditionally linear structure. For example, the Repressilator model, a popular CRN model (a more detailed introduction can be found in Section \ref{sec:stoch-Repressilator}), is a 6-dimensional system given by 
\begin{align}
   \label{eqP} \rd P_i &= \beta(M_i - P_i) \, \rd t \\
  \label{eqM}  \rd M_i & = (\alpha_0 + H(P_j) - M_i) \, \rd t, \quad j = (i + 1\mod 3) + 1
\end{align}
for $i = 1, 2, 3$, where $H(P_j) = \alpha K ^ n / (K^n + P_j^n)$ is the Hill function \cite{gesztelyi2012hill}. By looking at these equations, it is clear that the Repressilator exhibits conditional linearity, as  the ODEs \eqref{eqP} are linear, while those in \eqref{eqM} are conditionally linear in $M_i$ for fixed $P_j$. Moreover, it has also oscillatory behavior, so it is 
a compelling example for applying the splitting integrator of \cite{chen2020structure}. 
The preservation of the limit cycle by the splitting scheme is demonstrated in Figure \ref{fig:rk-vs-split-limitcycle}, bottom row, while the commonly used Runge-Kutta method fails in doing so for larger time steps.

\section{Data-conditional ABC-SMC}
\label{sec:appendix-dcabcsmc}
Recall $Y^{\rm DC} = Y^{\rm DC}_{1:d_o, 1:n}$.
\small
This appendix contains the ABC-SMC-DC Algorithm \ref{algo:sl-abs-smc} developed in \cite{jovanovski2023towards}.
\begin{algorithm}[t]
    \caption{Dynamic ABC-SMC with data-conditional simulation \textcolor{black}{(ABC-SMC-DC)}}\label{algo:sl-abs-smc}

    \begin{algorithmic}[1]
    \STATE Initialize the dataset $\mathcal{D}=\emptyset$.
    \STATE Generate $G$ i.i.d. prior-predictive samples $({Y}^{1:G}_{1:d_o}, \theta^{1:G}) \sim \Plik(Y) \, \pi(\theta)$ and store in $\mathcal{D}$.
    \STATE Train PEN on $\mathcal{D}$ to obtain $S_1(\cdot)$ and summarize the observation ${S}^{\rm o}_1 = S_1({Y}^{\rm o})$. 
    \FOR{$i = 1$ to $M$}
            \STATE Sample ${\theta}^i_{{1}} \sim \pi({\theta})$, run Algorithm \ref{algo:forward-path} to obtain $(\widetilde{Y}^{1:P}_{1:d_o, 1:n}, \omega^{1:P}_{1:n})$, and $Y^{\rm DC}$, and compute $d_1^i = \Vert S_1(Y^{\rm DC}) - {S}^{\rm o}_1 \Vert$.
            \STATE Compute the parameter weight $w_1^i$ by $\phi_p(S_r(Y^{\rm DC}) \given {\mu}_{P, {\theta}}, {\Sigma}_{P, {\theta}})/ \phi_p(S_r(Y^{\rm DC}) \given {{\mu}}^{\rm DC}_{P, {\theta}}, {{\Sigma}}^{\rm DC}_{P, {\theta}})$.
            \STATE Pick the trajectory ${Y}$ from the forward trajectories $\widetilde{Y}^{1:P}_{1:d_o, 1:n}$ that is closest (in Euclidean distance) to $\Obs$, and store in $\mathcal{D} \coloneqq \mathcal{D} \cup ({Y}, {\theta}^i_{{1}})$.
        \ENDFOR
        \FOR{$r = 2$ to $R$}
                \STATE Normalize $w_{r - 1}^{1:M}$.
                \STATE Retrain PEN on $\mathcal{D}$ to obtain $S_r(\cdot)$ and summarize ${S}^{\rm o}_r = S_r({Y}^{\rm o})$.
            \STATE Compute particle covariance ${\Sigma}_r = 2 \times {\rm Cov}((\bs{\theta}^{1:M}_{r - 1}, w_{r - 1}^{1:M}))$.
            \STATE Take $\epsilon_r$ to be the $\alpha$-quantile of distances $d_{r - 1}^{1:M}$ corresponding to accepted particles.
            \FOR{$i = 1$ to $M$}
                \WHILE{parameter not accepted}
                    \STATE Sample $\bs{\theta}^*$ from $\bs{\theta}_{r - 1}^{1:M}$ with probabilities $w_{r - 1}^{1:M}$ and perturb $\bs{\theta}^{i}_r \sim \mathcal{N}(\bs{\theta}^*, {\Sigma}_r)$. 
                    \STATE Run Algorithm \ref{algo:forward-path} to obtain $(\widetilde{Y}^{1:P}_{1:d_o, 1:n}, \omega^{1:P}_{1:n})$, and $Y^{\rm DC}$, and compute $d_r^i = \Vert S_r(Y^{\rm DC}) - {S}^{\rm o}_r \Vert$.
                    \IF{$d_r^i \leq \epsilon_r$} 
                 \STATE Accept ${\theta}^i_r$, compute the parameter weight $w_r^i$ by \eqref{eq:seq-imp-weight}.
                 \STATE Pick the trajectory ${Y}$ from the forward trajectories $\widetilde{Y}^{1:P}_{1:d_o, 1:n}$ that is closest (in Euclidean distance) to $\Obs$, and store in $\mathcal{D} \coloneqq \mathcal{D} \cup ({Y}, {\theta}^i_{{r}})$.
                        \ENDIF
                \ENDWHILE
            \ENDFOR
        \ENDFOR
        \STATE \textbf{Output:} Weighted sample $({\theta}^{1:M}_R, w_{R}^{1:M})$ of the ABC posterior distribution.
    \end{algorithmic}
\end{algorithm}


\section{Numerical splitting schemes for the considered CIR-type SDEs}
In this section, we apply the proposed numerical splitting scheme described in Section \ref{sec:cond-cir} to the three SDEs of interest, explicitly stating the chosen decomposition and deriving the solutions to the corresponding subequations.

\subsection{Stochastic Repressilator}\label{SectionC1}
When looking at the Repressilator SDE \eqref{repr}, we notice that the Brownian motions are not shared between the components (the mRNAs and the proteins), which corresponds to Scenario 2) in Section \ref{sec:subeq}). Hence, the solution to that system (in the weak sense) is also the solution to the following equivalent system
\begin{align}
    \rd M_i(t) &= \left( \alpha_0 + H(P_j(t)) - M_i(t) \right) \rd t + \sqrt{\alpha_0 + H(P_j(t)) + M_i(t)} \, \rd \widetilde{W}^{(i)}(t), \nonumber \\
    \rd P_i(t) &= \beta \left( M_i(t) - P_i(t) \right) \rd t + \sqrt{\beta (M_i(t) + P_i(t))} \, \rd \widetilde{W}^{(i + 3)}(t), \nonumber 
\end{align}
where $\widetilde{W}^{(1:6)}$ are standard uncorrelated Brownian motions. 
Thanks to this, the system does not need to be split before applying the Lamperti transform. Applying it to the mRNA SDEs $Z_i(t) = g(M_i(t)) = \sqrt{M_i + \alpha_0 + H(p_{j, k})}$, results in the transformed SDE
\begin{equation}
    \rd Z_i(t) = \left( \frac{\alpha_0 + H(p_{j, k}) - \frac{1}{8}}{Z_i(t)} - \frac{1}{2} Z_i(t) \right) \, \rd t + \frac{1}{2} \, \rd \widetilde{W}_i(t).
\end{equation}
The drift term forms a Bernoulli differential equation, with the known solution $Z^{(1)}_{i, k+1} = \Phi_h^{(M_i, 1)}(z_{i, k})$, where
\begin{equation}
    \Phi_h^{(M_i, 1)}(z_{i, k}) = \frac{1}{2} \sqrt{8(\alpha_0 + H(p_{j, k})) \left(1 - \frac{1}{e^h}\right) + \frac{4 z_{i, k}^2 + 1}{e^h} - 1}.
\end{equation}
The stochastic part has  solution $Z^{(2)}_{i, k+1} = \Phi_h^{(M_i, 2)}(z_{i, k}) = z_{i, k} + \Delta \widetilde{W}^{(i)}_{k+1}/2$, where $\Delta \widetilde{W}^{(i)}_{k+1} \sim \mathcal{N}(0, h)$. The complete solution to the mRNA SDE is obtained via Lie-Trotter composition followed by the application of the inverse mapping:
\begin{equation}
    M_{i, k+1} = \Phi_h^{(M_i)}(m_{i, k}) = \left( \left( \Phi_h^{(M_i, 2)} \circ \Phi_h^{(M_i, 1)} \right) (g(m_{i, k})) \right)^2 - \alpha_0 - H(p_{j, k}). \label{eq:rep-m-flow}
\end{equation}
A similar approach can be done for the protein SDEs by applying the Lamperti transform $Z_{i + 3}(t) = g(P_i(t)) = \sqrt{\beta(m_{i, k} + P_i(t))}$. Also in this case, the drift term of the transformed SDE forms an ODE with the known solution $Z_{i + 3, k + 1}^{(1)} = \Phi^{(P_i, 1)}(z_{i + 3, k})$, where
\begin{equation}
    \Phi^{(P_i)}_h(z_{i + 3, k}) = \frac{1}{2} \sqrt{\frac{\beta(1 - 8 m_{i, k}) + 4 z_{i + 3, k}^2}{e^{\beta h}} - \beta(1 - 8 m_{i, k})}.
\end{equation}
The stochastic part has solution $Z_{i + 3, k + 1}^{(2)} = \Phi_i^{(P_i, 2)}(z_{i + 3, k}) = z_{i + 3, k} + (\beta/2) \Delta \widetilde{W}^{(i+3)}_{k+1}$, where $\Delta \widetilde{W}^{(i+3)}_{k+1} \sim \mathcal{N}(0, h)$. The complete solution is given by:
\begin{equation}
    P_{i, k+1} = \Phi_h^{(P_i)}(p_{i, k}) = \beta^{-1}\left(\left( \left( \Phi_i^{(P_i, 2)} \circ \Phi_i^{(P_i, 1)} \right) (g(p_{i, k})) \right)^2 - \beta m_{i, k}\right). \label{eq:rep-p-flow}
\end{equation}
We split the vector field $f$ of the stochastic Repressilator as $
    f = f^{(M_1)} + f^{(M_2)} + f^{(M_3)} + f^{(P_1)} + f^{(P_2)} + f^{(P_3)}$. 
When the protein levels are fixed, the vector fields $f^{(M_1)}, f^{(M_2)},$ and $f^{(M_3)}$ commute, as each evolves independently. Defining $f^{(M_1, M_2, M_3)} = f^{(M_1)} + f^{(M_2)} + f^{(M_3)}$, we obtain $\Phi_h^{(M_1, M_2, M_3)} = \Phi_h^{(M_1)} \circ \Phi_h^{(M_2)} \circ \Phi_h^{(M_3)}$, where the flows $\Phi_h^{(M_i)}$ are as given in \eqref{eq:rep-m-flow}. Similarly, when mRNA levels are fixed, the vector fields $f^{(P_1)}, f^{(P_2)},$ and $f^{(P_3)}$ commute, allowing us to set $f^{(P_1, P_2, P_3)} = f^{(P_1)} + f^{(P_2)} + f^{(P_3)}$ and $\Phi_h^{(P_1, P_2, P_3)} = \Phi_h^{(P_1)} \circ \Phi_h^{(P_2)} \circ \Phi_h^{(P_3)}$, with flows $\Phi_h^{(P_i)}$ as specified in \eqref{eq:rep-p-flow}. Despite the Repressilator's 6-dimensional nature, we approximate $f$ using two main flows, composed using 
a Strang composition, leading to
\begin{align}
(M_{1, k + 1},& M_{2, k + 1}, M_{3, k + 1}, P_{1, k + 1}, P_{2, k + 1}, P_{3, k + 1}) =\nonumber  \\ & \left(    \Phi_{h/2}^{(M_1, M_2, M_3)} \circ \Phi_{h}^{(P_1, P_2, P_3)} \circ \Phi_{h/2}^{(M_1, M_2, M_3)}\right)(M_{1, k }, M_{2, k }, M_{3, k }, P_{1, k }, P_{2, k }, P_{3, k}).
\end{align}
\subsection{Stochastic Lotka-Volterra model}\label{SectionC2}
To apply our splitting scheme to this system, we first rewrite SDE  \eqref{eq:lv-sde1} for species $X_1$ as \eqref{eq:cle-cond-gen}. 
Assuming that the system has been simulated up to time $\tau_k$ with state $(x_{1, k}, x_{2, k})$, we take $\widetilde{a}_1 = 0$ and $\widetilde{b}_1 = -(\theta_1 - \theta_2 x_{2, k})$. Moreover, since all reactions include $X_1$, it follows that $R_{-1} = \emptyset$, meaning that the SDE is inherently of CIR-type, and there is no need to isolate that component. The diffusion coefficients are defined as $\widetilde{c}_{1, 1} = \sqrt{\theta_1}$ and $\widetilde{c}_{1, 2} = -\sqrt{\theta_2 x_{2, k}}$. Finally, the solution for $X_{1, k + 1}$ is given by substituting $\widetilde{a}_1$, $\widetilde{b}_1$, $\widetilde{c}_{1, 1}$, $\widetilde{c}_{1, 2}$ in \eqref{eq:ode-sol}-\eqref{eq:linear-sol} with $\Phi_h^{(X_1, 3)}(x_{1, k}) = x_{1, k}$, and composing them using Lie-Trotter. We can then transform the second component \eqref{eq:lv-sde2} similarly to the first one, taking $\widetilde{a}_2 = 0$ and $\widetilde{b}_2 = -(\theta_2 x_{1, k + 1} - \theta_3)$. This SDE is also of CIR-type, as there are no reactions that exclude $X_2$. The diffusion coefficients are $\widetilde{c}_{2, 2} = \sqrt{\theta_2 x_{1, k + 1}}$ and $\widetilde{c}_{2, 3} = - \sqrt{\theta_3}$. The solution for $X_{2, k + 1}$ is given by substituting these variables in \eqref{eq:ode-sol}-\eqref{eq:linear-sol} with $\Phi_h^{(X_2, 3)}(x_{2, k}) = x_{2, k}$, and composing them using Lie-Trotter. As the Lotka-Volterra model is two-dimensional, we split the vector field with two main flows $f^{(X_1)}$ and $f^{(X_2)}$ as
\begin{equation}
    f = f^{(X_1)} + f^{(X_2)},
\end{equation}
with the two flows on an arbitrary interval $[\tau_k, \tau_{k + 1}]$ of length $h$  given by $\Phi_h^{(X_1)}$ and $\Phi_h^{(X_2)}$, respectively. The flows are combined using 
a Strang composition, leading to
\begin{equation}
(X_{1,k+1}, X_{2, k + 1})=\left( \Phi_{h/2}^{(X_1)} \circ \Phi_{h}^{(X_2)} \circ \Phi_{h/2}^{(X_1)}\right)(X_{1,k}, X_{2, k }).
\end{equation}
\subsection{Two-pool model}\label{SectionC3}
As for the Lotka-Volterra model, we first need to transform the SDE \eqref{eq:twopool-1}-\eqref{eq:twopool-2} into the form specified in \eqref{eq:cle-cond-gen} to be use our splitting scheme. 
Assuming that the system has been simulated up to time $\tau_k$ with state $(x_{1, k}, x_{2, k})$, we take $\widetilde{a}_1 = \theta_4 x_{2, k}$ and $\widetilde{b}_1 = \theta_1 + \theta_3$. There are reactions that include and other that exclude $X_1$, leading to $R_1 = \{1, 3\}$ and $R_{-1} = \{4\}$. For the reactions in $R_1$, the diffusion coefficients are $\widetilde{c}_{1, 1} = -\sqrt{\theta_1}$ and $\widetilde{c}_{1, 3} = -\sqrt{\theta_3}$, and for the reaction in $R_{-1}$, the diffusion coefficient is $\widetilde{c}_{1, 4} = \sqrt{\theta_4 x_{2, k}}$. Finally, the solution for $X_{1, k + 1}$ is given by substituting $\widetilde{c}_{1, 4}$ in \eqref{eq:cle-split2-sol}, and $\widetilde{a}_1, \widetilde{b}_1, \widetilde{c}_{1, 1}, \widetilde{c}_{1, 3}$ in \eqref{eq:ode-sol}-\eqref{eq:linear-sol}, and composing the flows using Lie-Trotter. Given the solution of the first component, we can similarly transform the second component \eqref{eq:twopool-2}. We take $\widetilde{a}_2 = \theta_3 x_{1, k + 1}$ and $\widetilde{b}_2 = \theta_2 + \theta_4$. There reactions including/excluding $X_2$, leading to $R_2 = \{2, 4\}$ and $R_{-2} = \{3\}$. For reactions in $R_2$, the diffusion coefficients are $\widetilde{c}_{2, 2} = -\sqrt{\theta_2}$ and $\widetilde{c}_{2, 4} = - \sqrt{\theta_4}$, while for the reaction in $R_{-2}$, the diffusion coefficient is $\sqrt{\theta_3 x_{1, k + 1}}$. The solution for $X_{2, k + 1}$ is given by substituting $\widetilde{c}_{2, 3}$ in \eqref{eq:cle-split2-sol}, and $\widetilde{a}_2, \widetilde{b}_2, \widetilde{c}_{2, 2}, \widetilde{c}_{2, 4}$ in \eqref{eq:ode-sol}-\eqref{eq:linear-sol}, and composing the flows using Lie-Trotter. Similarly for the Lotka-Volterra model, the two-pool model is two-dimensional, so we split the vector field with two main flows $f^{(X_1)}$ and $f^{(X_2)}$ as
\begin{equation}
    f = f^{(X_1)} + f^{(X_2)},
\end{equation}
with the two flows on an arbitrary interval $[\tau_k, \tau_{k + 1}]$ of length $h$  given by $\Phi_h^{(X_1)}$ and $\Phi_h^{(X_2)}$, respectively. The flows are then combined using a Lie-Trotter composition, leading to
\begin{equation}
   (X_{1,k+1}, X_{2, k + 1})= \left(\Phi_{h}^{(X_2)} \circ \Phi_{h}^{(X_1)}\right)(X_{1,k}, X_{2, k }).
\end{equation}
\end{document}